%% file: main.tex
\documentclass[aps,prb,twocolumn, amsfonts, amssymb, reprint, showkeys, nofootinbib, superscriptaddress, twocolumn]{revtex4-2}

\usepackage[normalem]{ulem}
\usepackage{filecontents}
\usepackage[english]{babel}
\usepackage[utf8]{inputenc}
\usepackage{enumitem}
\usepackage[colorinlistoftodos, color=green!40, prependcaption]{todonotes}
\usepackage[pdftex, pdftitle={Article}, pdfauthor={Author}]{hyperref}
\newcommand{\be}{\begin{equation}}
\newcommand{\ee}{\end{equation}}
\newcommand{\ba}{\begin{eqnarray}}
\newcommand{\ea}{\end{eqnarray}}
\usepackage{upgreek}
\usepackage[tbtags]{amsmath}

\newcommand{\bigcircle}{%
  \raisebox{-0.2ex}{\scalebox{1.5}{$\bullet$}}%
}
\newcommand{\squareb}{
  \smash{\scalebox{0.8}{$\blacksquare$}}
  \vphantom{\blacksquare}
}

\begin{document}
\title{Resonant Magneto-phonon Emission by Supersonic Electrons in Ultra-high Mobility Two-dimensional System}
 
\author{Z. T.~Wang}
\author{M.~Hilke}
\email[]{michael.hilke@mcgill.ca}
\affiliation{Department of Physics, McGill University, Montr\'eal, Quebec, Canada, H3A 2T8}
\author{N.~Fong}
\affiliation{Emerging Technology Division, National Research Council of Canada, Ottawa, Ontario, Canada, K1A 0R6}
\author{D.~G.~Austing}
\email[]{guy.austing@nrc-cnrc.gc.ca}
\affiliation{Department of Physics, McGill University, Montr\'eal, Quebec, Canada, H3A 2T8}
\affiliation{Emerging Technology Division, National Research Council of Canada, Ottawa, Ontario, Canada, K1A 0R6}
\author{S.~A.~Studenikin}
\affiliation{Emerging Technology Division, National Research Council of Canada, Ottawa, Ontario, Canada, K1A 0R6}
\author{K. W.~West}
\author{L. N.~Pfeiffer}
\affiliation{Department of Electrical Engineering, Princeton University, Princeton, New Jersey, USA, 08544}


\begin{abstract}
We investigate resonant acoustic phonon scattering in the magneto-resistivity of an ultra-high mobility two-dimensional electron gas system subject to DC current in the temperature range 10 mK to 3.9 K. For a DC current density of $\sim$1.1 A/m, the induced carrier drift velocity $v_{drift}$ becomes equal to the speed of sound $s \sim$ 3 km/s. When $v_{drift} \gtrsim s$ very strong resonant features with only weak temperature dependence are observed and identified as phonon-induced resistance oscillations at and above the ``sound barrier''. Their behavior contrasts with that in the subsonic regime ($v_{drift} < s$) where resonant acoustic phonon scattering is strongly suppressed when the temperature is reduced unless amplified with quasi-elastic inter-Landau-level scattering. Our observations are compared to recent theoretical predictions from which we can extract a dimensionless electron-phonon coupling constant of $g^{2}$=0.0016 for the strong non-linear transport regime. We find evidence for a predicted oscillation phase change ' effect on traversing the ``sound barrier''. Crossing the “sound barrier” fundamentally alters the resulting phonon emission processes, and the applied magnetic field results in pronounced and sharp resonant phonon emission due to Landau level quantization.
\end{abstract}

\maketitle

 
Evidence for emission of acoustic phonons by supersonic carriers when under high electric field the carrier drift velocity $v_{drift}$ exceeds the sound velocity $s$ in metal and semiconductor crystals was described over half a century ago \cite{kaganov1957,Hutson1961,Esaki1962,Smith1962,Wang1962,Spector1963,Eckstein1963,Prohofsky1964,Zylber1967}. Such work has motivated the recent drive to exploit sound amplification for efficient and intense on-chip electrically-driven sources of coherent acoustic photons, i.e., a phonon laser [see for example Refs. \cite{Beardsley2010, Maryam2013, Shinokita2016, Barajas2024, Wendt2025}], and inspired a broad range of experiments that reveal numerous phenomena that can be explained by analogy to hydrodynamic shock waves (``sonic boom''), the Mach cone, Cherenkov radiation, and plasmon-Cherenkov effects \cite{kaganov1957,Abajo2010,Shinokita2016,andersen2019electron,greenaway2021,Hu2022,Barajas2024,Geurs2025,dong2025current}.

In a modern context, magneto-phonon resonances of carriers confined in two-dimensions under equilibrium and non-equilibrium conditions when subjected to a weak or moderate strength out-of-plane ($B$-) magnetic field have attracted considerable attention \cite{vitkalov2009NL, dmitriev2011, dmitriev2012review}. Initial work focused on phonon-induced resistance oscillations (PIROs) observed under equilibrium conditions (no DC current applied) in a field up to a few hundred mT and typically in a temperature range of $\sim$2 to 10-20 K \cite{epoint} that are periodic in 1/$B$ \cite{zudov2001, yang2002A, bykov2005MP, Zhang2004XYZ, hatke2009phonon, Bykov2009, bykov2010, Hatke2010, hatke2011piro, hatke2012giant, kumara2019, Greenaway2019}. The effect is explained in terms carrier scattering between Landau levels (LLs) through \emph{resonant} absorption or emission of acoustic phonons accompanied by momentum transfer $\sim$2$k_F$, equivalent to a displacement by $\sim$2$R_c$ of the carriers' guiding center perpendicular to the Hall bar channel, when the corresponding phonon energy $\sim$$2\hbar k_F s$ bridges integer multiples of the LL energy separation (the cyclotron energy $\hbar\omega_c$), i.e., $\epsilon_{ph} = 2k_{F} s/ \omega_c$ $\approx  p = \pm$1, $\pm$2, $\pm$3, $\ldots$ \cite{zudov2001,zhang2008}. Here $k_F$, $R_c$, $\hbar$ and $\omega_c$ are respectively the carrier Fermi wave number, the cyclotron radius, the reduced Planck's constant and the cyclotron frequency. 

\begin{figure*}[]
\includegraphics[width=2\columnwidth] {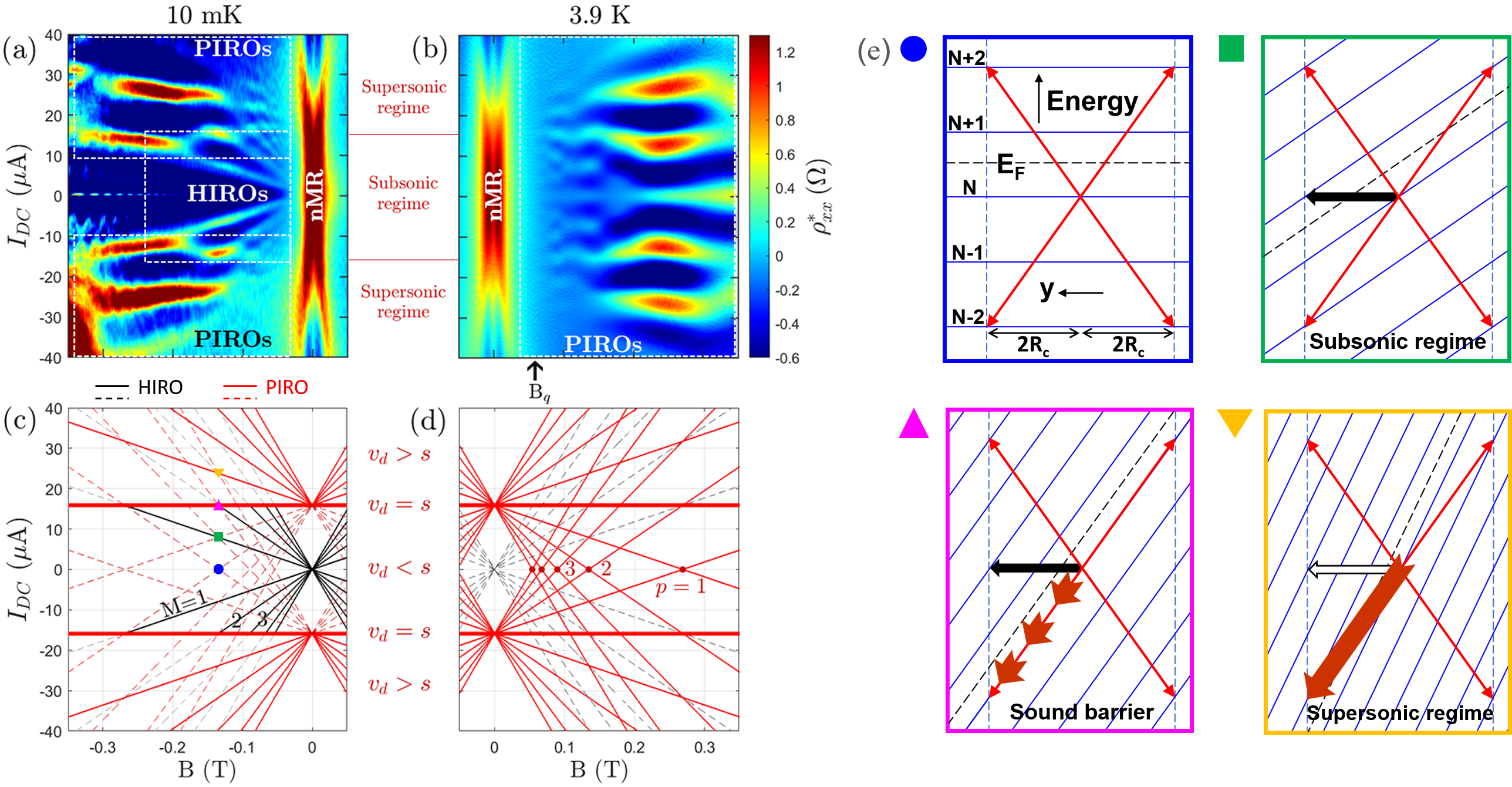}
\caption{
$\rho^{*}_{xx}(B,I_{DC})$ at (a) 10 mK for $-B$, and (b) 3.9 K for $+B$. See Fig. \ref{fig:bigfig} for full plots of $\rho^{*}_{xx}$ at 10 mK and 3.9 K and Appendix C for data measured at intermediate temperature 1.2 K. The characteristic $B$-field $B_{q} =$ 60 mT for the onset of PIROs is marked in (b)- see discussion in SM \cite{supplemental} Sec. S4. Plots showing expected path of HIROs (black lines) and PIROs (red lines) at (c) 10 mK for $-B$, and (d) 3.9 K for $+B$, and their expected relative amplitude: solid (dashed) lines indicate HIRO and PIRO features anticipated to be observable (weak or absent). See SM \cite{supplemental} Sec. S2 (Sec. S3) for details of how the black HIRO (red PIRO) lines are calculated. HIRO lines with order $M=\pm$1, $\pm$2, $\pm$3, $\pm$4, $\pm$5 are shown. PIRO lines converging to points at zero DC current that can give rise to PIRO peaks under equilibrium conditions with index $p= \pm$1, $\pm$2, $\pm$3, $\pm$4, $\pm$5 are shown. See Appendix D for comments related to the sign convention adopted for $M$ and $p$. The ``sound barrier'' condition $v_{drift} = s$ is indicated by the bold red line. (e) Cartoons of LLs (blue lines indexed $\ldots$, $N-2, N-1, N, N+1, N+2$, $\ldots$) near the Fermi level (black dashed line labeled $E_{F}$) depicting various possible (exemplary) resonant scattering processes involving acoustic phonons that can lead to PIROs (red arrows) and a quasi-elastic scattering process that can lead to HIROs (black arrow). The four cartoons show possible allowed processes at the four points indicated by the colored symbols in (c), all at the same $B$-field. See text, Appendix E and SM \cite{supplemental} Sec. S3 for full discussion.}
\label{fig:summary}
\end{figure*} 

Subsequent investigations considered PIROs under non-equilibrium conditions by application of a DC current \cite{zhang2008}. A DC current $I_{DC} = neWv_{drift}$ not only regulates $v_{drift}$ but induces a Hall electric field ($= v_{drift}B$) that tilts the LLs along a direction (y) perpendicular to the current flow along the Hall bar channel. Here $n$ ($= k_F^{2}/2\pi$), $e$, and $W$ respectively are the two-dimensional (2D) carrier density, charge of an electron, and Hall bar width. Zhang \textit{et al.} \cite{zhang2008} outlined three regimes [Fig. \ref{fig:summary}(e)]: i. a subsonic regime ($v_{drift} < s$) at low $I_{DC}$ (weakly tilted LLs) where \emph{inter}-LL scattering of carriers can occur through resonant phonon absorption and emission that is limited at low temperature respectively by the vanishing ambient thermal phonon population and by lack of empty electronic states below the Fermi energy; ii. a critical point where $v_{drift} = s$ (the ``sound barrier'') at higher $I_{DC}$ when momentarily \emph{intra}-LL scattering is also possible by emission of phonons with any wave-vector up to $\sim$2$k_F$; and iii. a supersonic regime ($v_{drift} > s$) at yet higher $I_{DC}$ (strongly tilted LLs) where $\emph{inter}$-LL scattering of carriers from a lower (occupied) LL to a \emph{higher} (empty) LL accompanied by resonant emission of phonons can now freely occur- a process that is strong even at very low temperature \cite{dmitriev2010}. PIROs in the presence of a DC current applied to high mobility 2D carrier systems in GaAs/AlGaAs (for electrons) and in graphene (for holes) have been reported for $v_{drift}$ approaching $s$ but not beyond \cite{zhang2008,greenaway2021, hatke2015SAND, raichev2017, dpoint}. While a theory for PIROs based on a semi-classical picture featuring cyclotron orbits encompassing both the subsonic and supersonic regimes has been developed \cite{dmitriev2010}, it has yet to be tested against experimental data. This theory specifically predicts strong PIROs in the supersonic regime whose amplitude saturates with lowering temperature.

Here, for temperature $T<$ 4 K we study non-linear transport of a narrow Hall bar fabricated from an ultra-high mobility GaAs/AlGaAs 2D electron gas (2DEG) hetero-structure, with an applied DC current density up to $\sim$$2.7$ A/m. Compared to previous studies, the applied DC current density is sufficiently high to ensure that $v_{drift}$ can exceed $s$ by a factor of $\sim$2.5. This enables us to explore deep into the supersonic regime where we observe pronounced PIROs arising from strong resonant phonon emission even at milli-Kelvin temperatures. Experimental details are provided in Appendix A.  

Figure \ref{fig:summary} presents plots of the longitudinal differential resistivity as a function of $B$ and $I_{DC}$ measured at (a) 10 mK and (b) 3.9 K. To focus on the resonant features of interest, we henceforth discuss $\rho^{*}_{xx}$: the measured differential resistivity $\rho_{xx}$ minus a smooth background- see Supplemental Material (SM) \cite{supplemental} Sec. S1. At both 10 mK and 3.9 K, within a couple tens of milli-Tesla of zero field, and extending up to high DC current, we observe large negative magneto-resistance (nMR) \cite{mani2013size,Bockhorn2011,hatke2012giant,Bockhorn2013,shi2014,Shi2014c,Bockhorn2014,Schluck2015,Wang2016,Samar2017,Samar2018,Schluck2018,Samar2020,HornCosfeld2021,Wang2022,paper1, Bockhorn2024,Bartels2025} and a double-peak feature \cite{paper1}, as well as Shubnikov de Haas (SdH) oscillations clearest at 10 mK at zero DC current.

Our focus here is on phenomena occurring at a weak to moderate strength $B$-field beyond the nMR region and DC currents up to 40 \textmu A. In the region $|I_{DC}| \lesssim$ 15 \textmu A, at 10 mK in Fig. \ref{fig:summary}(a), we observe Hall field-induced resistance oscillations (HIROs) that arise from quasi-elastic inter-LL scattering \cite{yang2002zener,Bykov2005,zhang2007MT,zhang2007effect,vavilov2007,Lei2007,bykov2008,Hatke2009,vitkalov2009NL,Hatke2011,dmitriev2012review,Hatke2012,Shi2014b,hatke2015SAND,shi2017,Zudov2017,greenaway2021,paper1,Bartels2025}. A distinguishing feature of the HIROs is that they fan out in lines from the origin. In Sec. S2 of SM \cite{supplemental} we fully characterize the HIROs. The HIROs rapidly wash out beyond $\sim$$15$ $\mu$A. The (almost horizontal-running) resonance observed at this current and the resonances beyond have a different origin- they are PIROs. The strong resonances centered at -0.25 T located at approximately $\pm15$ $\mu$A and $\pm30$ $\mu$A are part of a string of features of different amplitude that track to common points at finite DC current at zero field. Fainter but nevertheless discernible resonances that track to the common points can also be observed, especially in the $+I_{DC}$ quadrant, at higher DC current. We attribute the boundary near $15$ \textmu A to the ``sound barrier'', and identify the region to higher (down to zero) current as the supersonic (subsonic) regime. From detailed analysis in Appendix B we determine $v_{drift} = s =$ 3.0 km/s (equivalent to $I_{DC} = \pm 15.9$ \textmu A). At 10 mK, HIROs (PIROs) inhabit the subsonic (supersonic) regime. On the other hand, at 3.9 K in Fig. \ref{fig:summary}(b), the situation is different. While the PIROs in the supersonic regime have changed comparatively little (see the four strong resonances centered at +0.25 T), the distinctive HIROs are no longer visible in the subsonic regime. Instead, a different pattern of features (array of spot-like resonances) is present that are PIROs in origin \cite{zhang2008,greenaway2021}, i.e, at this elevated temperature PIROs are dominant in both the subsonic and supersonic regimes.


\begin{figure}[] 
\includegraphics[width=1.0\columnwidth]{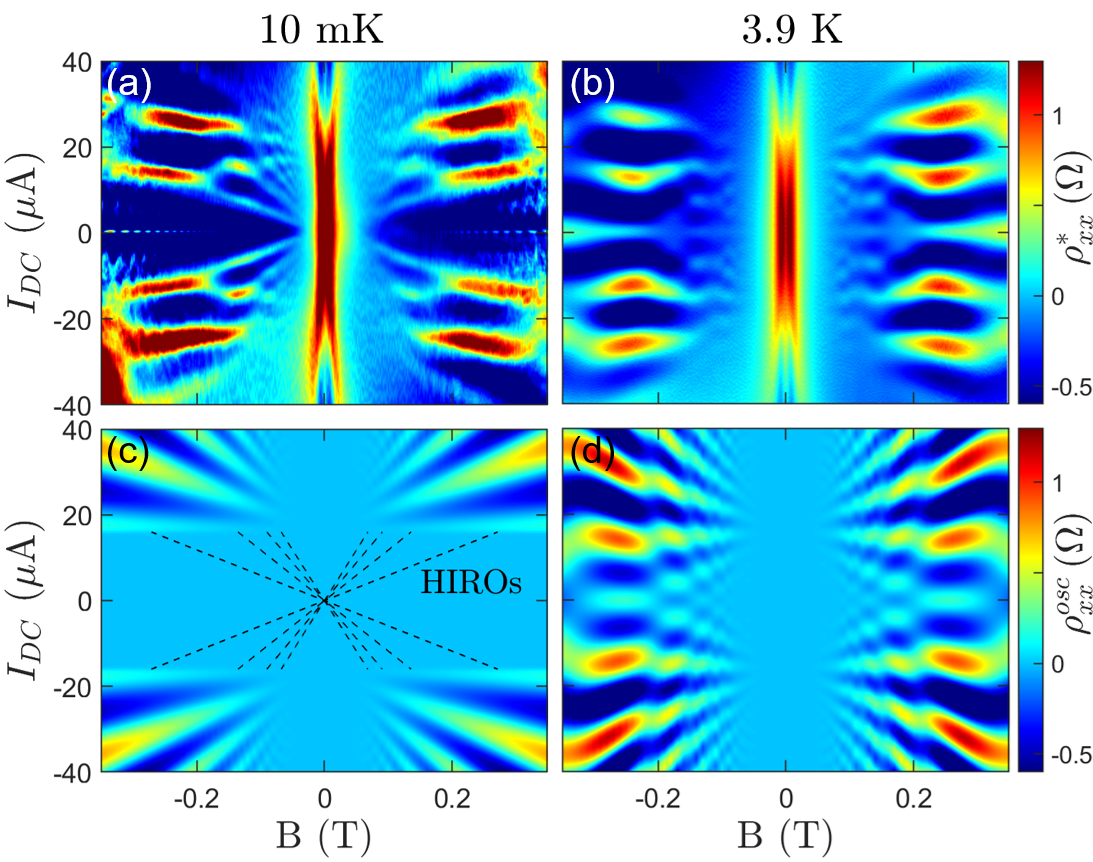}
\caption{Plots of $\rho^{*}_{xx}$ for (a) 10 mK and (b) 3.9 K. Corresponding plots (c-d) of oscillatory component of PIROs $\rho^{osc}_{xx}$ calculated from model in \cite{dmitriev2010}. See discussion in text and SM \cite{supplemental} Sec. S4 regarding parameters used in the calculations. For data in (a), the resonant lines radiating out from the origin in the subsonic regime ($|I_{DC}|<15.9$ \textmu A) are HIROs. Quasi-elastic inter-LL scattering  responsible for HIROs is not incorporated in the model described in \cite{dmitriev2010}. For reference, we have added the black dashed lines in the subsonic regime in (c) that track the calculated position of the HIRO features extracted from the data in (a)- see SM \cite{supplemental} Sec. S2.}
\label{fig:bigfig}
\end{figure} 

As a guide to interpreting the experimental data, Fig. \ref{fig:summary} shows plots marking the expected position from calculation of HIROs (black lines: see SM \cite{supplemental} Sec. S2) and PIROs (red lines: see SM \cite{supplemental} Sec. S3) at (c) the lowest temperature for $-B$, and at (d) the elevated temperature for $+B$. Solid (dashed) lines identify features anticipated to be observable (weak or absent). At zero field, all PIRO lines converge to two common points at $I_{DC} = \pm$ 15.9 $\mu$A set by the determined value of the sound velocity $s =$ 3.0 km/s, and all HIRO lines converge to zero DC current. For illustrative purposes, Fig. \ref{fig:summary}(e) depicts certain phonon absorption and emission scattering processes between and within LLs that can lead to PIROs (red arrows: see SM \cite{supplemental} Sec. S3 for further details) and a quasi-elastic scattering process between LLs that can lead to HIROs (black arrow). The cartoons are pertinent to the situations at the four colored symbols marked in Fig. \ref{fig:summary}(c) as canonical examples. The possible allowed process give rise to: i. the conventional second-order PIRO peak at zero DC current at sufficiently high temperature (blue $\bigcircle$: LLs un-tilted); ii. the first-order HIRO peak at finite current in the subsonic regime at low temperature enhanced by a crossing of two PIRO lines, or a PIRO peak arising from just the crossing of the two PIRO lines at higher temperature (green $\squareb$: LLs tilted by Hall field along the y-direction); iii. the second-order HIRO peak at higher current at low temperature coinciding with the red bold line in (c) at the ``sound barrier'' along which \emph{intra}-LL scattering by emission of phonons with any wave-vector up to $\sim$2$k_F$ (short thick red arrows) can readily occur even at the lowest temperature (magenta $\blacktriangle$: LLs tilted further); and iv. the strong PIRO peak at yet higher current in the supersonic regime where one of the contributing processes involves resonant inter-LL scattering of electrons by emission of $\sim$2$k_F$ phonon (long thick red arrow) from a state in a largely occupied LL to a state in a largely unoccupied higher-index LL at \emph{lower energy}- a process that also can readily occur even at the lowest temperature (orange $\blacktriangledown$: highly tilted LLs).


\begin{figure*}[]
\hspace*{-0.05in}
\includegraphics[width=1.8\columnwidth]{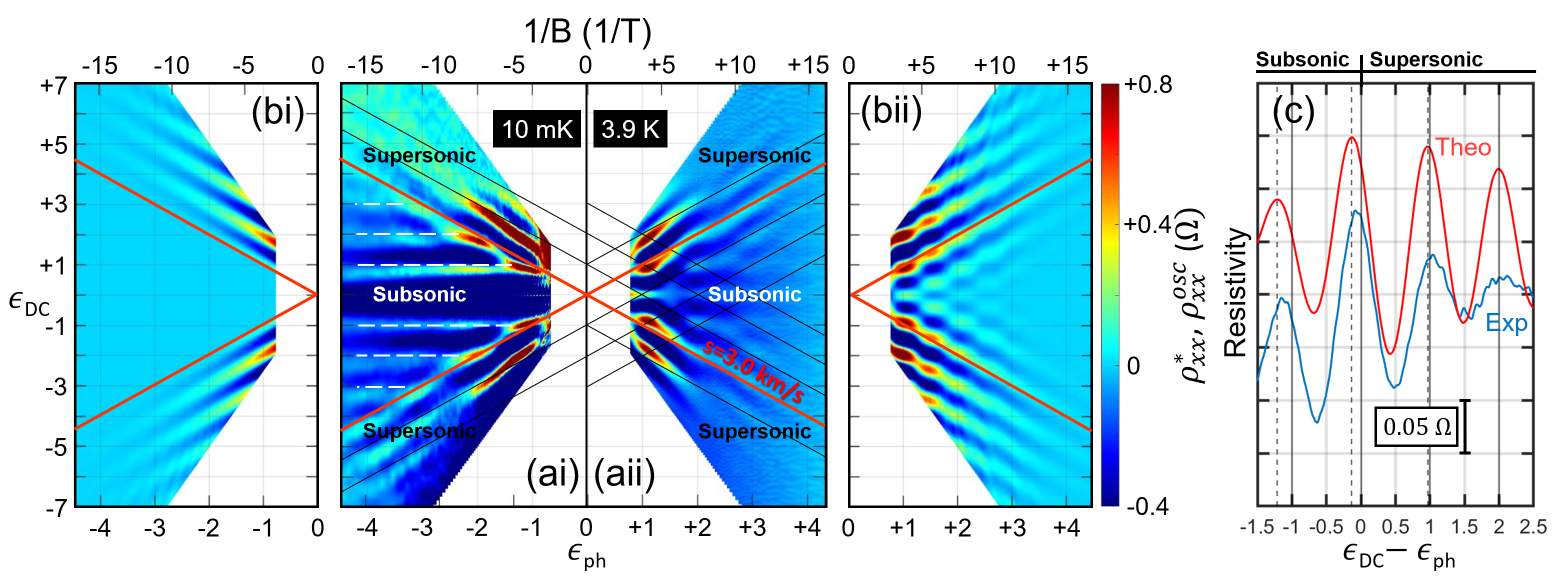} 
\caption{$\rho^{*}_{xx}$ at (ai) 10 mK for $-B$, and (aii) 3.9 K for $+B$, plotted as a function of $\epsilon_{ph}$ and $\epsilon_{DC}$ following \cite{dmitriev2010}. The subsonic and supersonic regimes separated by the ``sound barrier'' for $s=3.0$ km/s are identified. Some black diagonal and anti-diagonal lines with integer value of $\epsilon_{DC}-\epsilon_{ph}$ acting as guide lines for anticipated PIRO line features are included above the ``sound barrier'' in (ai) and either side of the ``sound barrier'' in (aii). Dashed white lines for $\epsilon_{DC}= \pm1, \pm2, \pm3$ highlight HIROs up to third order in the 10 mK data in (ai). Corresponding plots (bi) and (bii) of $\rho^{osc}_{xx}$ calculated from the model \cite{dmitriev2010}. See SM \cite{supplemental} Sec. S5 for full discussion regarding interpretation of transformed data and predicted $\pi/2$ phase change effect on transitioning from the subsonic to the supersonic regime \cite{dmitriev2010}. Compare theoretical red trace (Theo) with experimental blue trace (Exp) in (c) as evidence for the effect: see the PIRO peak in the supersonic [subsonic] regime located near $\epsilon_{DC}-\epsilon_{ph}= +1$ $[= -(1 + \frac{1}{4})]$.} 
\label{fig:edceph}
\end{figure*}


Next, we examine the evolution of $\rho^{*}_{xx}$ with temperature and compare with calculations based on the theory of Dmitriev \textit{et al.} \cite{dmitriev2010}. Figure \ref{fig:bigfig} shows $\rho^{*}_{xx}$ for (a) 10 mK and (b) 3.9 K. Here we show the full plots for 10 mK and 3.9 K rather than the half plots in Figs. \ref{fig:summary}(a) and \ref{fig:summary}(b) confirming the high symmetry of the data with respect to the $B$-field. Over the 10 mK to 3.9 K range, the most striking aspects of the data plots are the pronounced PIRO resonances in the supersonic regime ($|I_{DC}|>15.9$ $\mu$A) with a comparatively weak temperature dependence, and the emergence of the strongly temperature-dependent PIROs in the subsonic regime at 3.9 K. Also, while the HIRO fan is clear in the subsonic regime in the 10 mK data, albeit with modulations in amplitude when a HIRO line intersects expected crossings of PIRO lines (the ``double resonance'' effect \cite{zhang2008}), HIROs are rapidly diminished on entering the supersonic regime, and in the 3.9 K data HIROs are absent everywhere (see SM \cite{supplemental} Sec. S2 for further discussion). See Appendix C for data at 1.2 K that highlights the transition in the subsonic regime from HIRO-dominated at 10 mK to PIRO-dominated at 3.9 K.

We now consider whether our observations regarding PIROs are consistent with the predictions of the model in \cite{dmitriev2010}. Aside from features in the experimental data related to SdH oscillations, nMR, and HIROs, which are not incorporated in the model, we assume features in $\rho_{xx}^{*}$ that we attribute to PIROs can be correlated directly with features in the calculated electron-phonon transport scattering rate- see SM \cite{supplemental} Secs. S1 and S4. Figure \ref{fig:bigfig} bottom row shows plots of the oscillatory component of PIROs $\rho^{osc}_{xx}$ calculated from the model to compare with the data plots in the top row. The parameters used are $s=3.0$ km/s, the electron-phonon coupling constant $g^2=0.0016$, and the quantum lifetime $\tau_q=6.3$ ps [estimated from the characteristic field $B_{q}$ marked in Fig. \ref{fig:summary}(b)]- see SM \cite{supplemental} Sec. S4 for discussion on the value of $g^2$. For the 10 mK data, the temperature used for the calculation is taken to be the estimated 40 mK electron temperature rather than the 10 mK mixing chamber temperature- see Appendix F. We find qualitative agreement between the $\rho_{xx}^{*}$ data and $\rho^{osc}_{xx}$ calculations. Namely, in the subsonic regime, the model calculations reveal that PIROs are strongly suppressed at very low temperature but grow quickly in amplitude with increasing temperature, whereas in the supersonic regime, the model calculations reveal that the amplitude of the PIROs saturates at very low temperature. Nonetheless, on a fine scale there are clearly differences in details in the supersonic regime. In the data, compared to the calculations: the pronounced (red colored) PIRO lobes near $\pm$30 $\mu$A at $\pm$250 mT are ``flatter'', and are surprisingly stronger at 10 mK than at 3.9 K by a factor of approximately two (see SM \cite{supplemental} Sec. S3); the PIROs wash out more rapidly approaching $\pm$40 $\mu$A; and the nature of the modulated PIRO ``wiggles'' in amplitude and position is different most notably in the 10 mK data. We note that electron heating is a factor when comparing our data to the theory in \cite{dmitriev2010} where electron heating is not included- see Appendix F.

Lastly, we transform the 10 mK and 3.9 K data measured on sweeping $I_{DC}$ and stepping $B$ to the $(\epsilon_{ph},\epsilon_{DC})$ coordinate system where $\epsilon_{DC} = 2I_{DC}k_F/neW\omega_c$ \cite{dmitriev2010}- see half plots (ai) and (aii) [and corresponding calculated plots (bi) and (bii)], respectively, in Fig. \ref{fig:edceph}. See Appendix D for comments related to the sign convention adopted for $\epsilon_{DC}$ and $\epsilon_{ph}$. Plotted this way, PIRO line features track diagonal and anti-diagonal lines with integer value of $\epsilon_{DC} \pm \epsilon_{ph}$ [see selected black and red solid guide lines in (ai) and (aii)] and HIRO lines features map to horizontal lines with integer values of $\epsilon_{DC}$ [see selected white dashed lines in (ai)]. See SM \cite{supplemental} Sec. S5 for further details of the transform relating to PIROs. Dmitriev \textit{et al.} \cite{dmitriev2010} predicted a subtle effect- a progressive $\pi/2$ change in the PIRO phase near the ``sound barrier'' on transitioning from the subsonic regime to the supersonic regime. Either side of the ``sound barrier'' $\epsilon_{DC}-\epsilon_{ph}=0$, in the supersonic [subsonic] regime, PIRO maxima are expected to occur at or close to $\epsilon_{DC}-\epsilon_{ph}= +1, +2, ...$ $[= ..., -(2+\frac{1}{4}), -(1+\frac{1}{4})]$. See traces in (c) as evidence for this effect in our 3.9 K data which we discuss thoroughly in SM \cite{supplemental} Sec. S5.


In conclusion, we have pushed observation of resonant acoustic phonon scattering under non-equilibrium conditions deep into the supersonic regime. A remarkable consequence is the dominance of electron-phonon interaction in this regime, even at very low temperature, where the ambient phonon population is low but the electron-induced phonon emission is nonetheless strong. Although the theory of PIROs in \cite{dmitriev2010} covering both the subsonic and supersonic regimes as well as a predicted $\pi/2$ phase change in PIROs on traversing the ``sound barrier'' qualitatively reproduces key features in the experimental data, we see subtle differences which invite theoretical attention. In particular, a theory incorporating both HIROs and PIROs would be useful (since it would shed light on the competition between HIRO and PIRO channels especially at the ``sound barrier'' and beyond where HIROs are strongly suppressed). Additionally, a theory addressing the full impact on the electron temperature of an applied DC current (electron heating) is desirable. A surprising feature of the 10 mK data is the significant (approximately a factor of two) enhancement of the PIRO amplitude in the supersonic regime at the lowest temperature not expected in the theory in \cite{dmitriev2010}. This points to a highly efficient low temperature acoustic phonon emission mechanism with a narrow frequency line, and raises the prospect of a magneto-phonon laser by incorporating an ultra-high mobility 2DEG in an acoustic resonator. A surface acoustic wave phonon laser with a 2DEG gain medium was recently demonstrated in \cite{Wendt2025} at room temperature and zero $B$-field. Our observations are applicable to carriers in any high mobility 2D system passing a sufficiently high DC current and the technique can shed further light on electron-phonon and hole-phonon coupling mechanisms.


We thank Mark Greenaway, Laurence Eaves, Lina Bockhorn and Rolf Haug for useful discussions. The authors acknowledge support from INTRIQ and RQMP. This research is funded in part by FRQNT, NSERC and the Gordon and Betty Moore Foundation’s EPiQS Initiative, Grant GBMF9615 to L. N. Pfeiffer, and by the National Science Foundation MRSEC grant DMR 2011750 to Princeton University.

\input{biblio.tex}

\onecolumngrid
\vspace{1\baselineskip}
\begin{center}
  {\Large \bfseries End Matter}
\end{center}
\twocolumngrid
\textit{Appendix A: Experimental details}---Measurements are made on a Hall bar with lithographic width $W=$ 15 \textmu m and adjacent side probes separated by a distance of 50 \textmu m. The Hall bar is fabricated from a GaAs/AlGaAs quantum well (QW) hetero-structure. The QW region consists of a 34.5 nm wide GaAs layer cladded with 60 nm wide layers of Al$_{0.12}$Ga$_{0.88}$As. The barriers are composed of Al$_{0.24}$Ga$_{0.76}$As layers which contain Si delta-doping set back $\sim$70 nm either side of the QW. For the Hall bar measured, after illumination, the 2DEG carrier density and mobility respectively are determined to be $n = (2.2 \pm 0.1) \times$ 10$^{11}$ cm$^{-2}$ and $\mu = (18 \pm 1) \times$ 10$^6$ cm$^2$/(Vs) at $\sim$10 mK. We note that in comparison, for a large area Van der Pauw device measured in a helium-3 cryostat at base temperature, the 2DEG carrier density and mobility respectively were determined to be $n$ = 2.04 $\times$ 10$^{11}$ cm$^{-2}$ and $\mu=$ 29.8 $\times$ 10$^6$ cm$^2$/(Vs) after illumination- see Ref. \cite{paper1} for commentary on mobility determination in narrow Hall bars. A DC current $I_{DC}$ of up to 40 \textmu A combined with an AC current of 200 nA or 400 nA at $\sim$30 Hz is driven along the Hall bar. The differential longitudinal resistivity $\rho_{xx}$ is measured directly using a standard lock-in technique (see \cite{phaseinversionYu,paper1} for further details). The $B$-field applied is out-of-plane with respect to the 2DEG. The experiments reported are performed in a dilution refrigerator at three different mixing chamber (bath) temperatures: $T$ $\sim$ 10 mK, 1.2 K and 3.9 K.

\begin{figure}[]
\includegraphics[width=0.8\columnwidth]{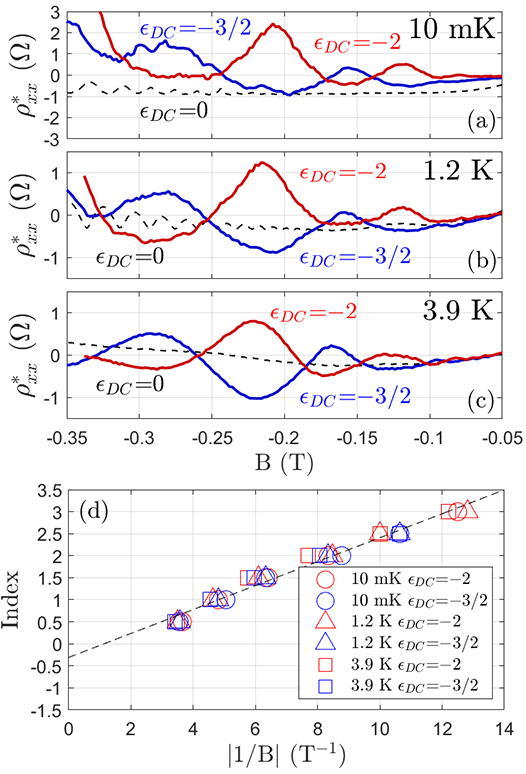}
\caption{$\rho^{*}_{xx}$ versus B traces for $\epsilon_{DC} =$ 0 (black dashed), $\epsilon_{DC}= $ $-3/2$ (blue) and $\epsilon_{DC}= $ $-2$ (red), at (a) 10 mK, (b) 1.2 K and (c) 3.9 K. Points are taken from data in the $+I_{DC}$ and $-B$ quadrant. (d) Position of successive extrema of oscillations in (a)-(c) for $\epsilon_{DC}=
$ $-2$, $-3/2$ plotted versus 1/B. The y-axis index-label 1, 2, 3 (0.5, 1.5, 2.5) for the maxima (minima) here effectively corresponding to $\epsilon_{ph}$. Note that for $\epsilon_{DC}=$ 0 (zero DC current), even at 3.9 K, PIROs extrema are not clearly resolved. The weak oscillations that develop with $B$ in the 10 mK and 1.2 K traces are SdH oscillations.}
\label{fig:vsound}
\end{figure} 

\textit{Appendix B: Sound velocity}---In Fig. \ref{fig:vsound}, we plot $\rho^{*}_{xx}$ versus $B$ at $\epsilon_{DC}= $ 0, $-3/2$, and $-2$ for (a) 10 mK, (b) 1.2 K and (c) 3.9 K. Here $\epsilon_{DC} = 2I_{DC}k_F/neW\omega_c$ and lines of constant integer value ($M$) for $\epsilon_{DC}$ trace out the anticipated HIRO lines (see SM \cite{supplemental} Sec. S2) \cite{yang2002zener,zhang2008}. The standard method \cite{zudov2001} to determine $s$ using extrema in PIROs at \emph{zero DC current} cannot be employed in our case because the PIRO amplitude is too weak over the temperature range available (see black dashed traces at $\epsilon_{DC}= $ 0). Instead, following Zhang \textit{et al.} \cite{zhang2008} we examine the oscillations in $\rho^{*}_{xx}$ that are visible for finite DC current along lines of constant integer and half-integer values of $\epsilon_{DC}$ and estimate $s$ from the position of the extrema. Here we shall only consider the position of the maxima and minima for the integer (half-integer) $\epsilon_{DC}$ value of $-2$ ($-3/2$). The $\epsilon_{DC}= $ $-2$ ($\epsilon_{DC}= $ $-3/2$) traces correspond to sections in $B$ versus $I_{DC}$ plots tracking the anticipated $M= $ $-2$ HIRO line (a line mid-way between the anticipated $M= $ $-1$ and $M= $ $-2$ HIRO lines). As expected, maxima in the $\epsilon_{DC}= $ $-2$ traces occur where there are minima in the $\epsilon_{DC}= $ $-3/2$ traces and vice versa. At 10 mK, when PIROs are otherwise strongly suppressed in the subsonic regime, the oscillation maxima in $\rho^{*}_{xx}$ at integer values of $\epsilon_{DC}$ arise when a crossing of PIRO lines coincides with a HIRO line [see Figs. \ref{fig:summary}(a) and \ref{fig:summary}(c)], i.e., the PIRO signal is essentially amplified by the strong HIRO signal- the ``double resonance'' effect \cite{zhang2008}. However, at 3.9 K, the oscillation maxima arise naturally just from the crossings of the PIRO lines, even when the HIRO signal is suppressed [see Figs. \ref{fig:summary}(b) and \ref{fig:summary}(d)]. Using the data points for all three temperatures, $s$ is found to be $3.0 \pm 0.2$ km/s from the slope [see Fig. \ref{fig:vsound}(d)]. See SM \cite{supplemental} Sec. S3 for discussion on the value of $s$ inferred. Across the three temperatures, the data points essentially lie on a common line, i.e., $s$ is approximately constant over the 10 mK- 3.9 K range. For $s=$ 3.0 km/s, from the expression $I_{DC}/W = nes$, the transition from the subsonic regime to the supersonic regime is expected to occur at $I_{DC} = \pm 15.9$ \textmu A. This current value is in close agreement with the position of the strong almost horizontal-running features located at $\sim\pm 15$ \textmu A in the plots in Figs. \ref{fig:summary}(a) and \ref{fig:summary}(b).

\begin{figure}[] 
\includegraphics[width=0.6\columnwidth]{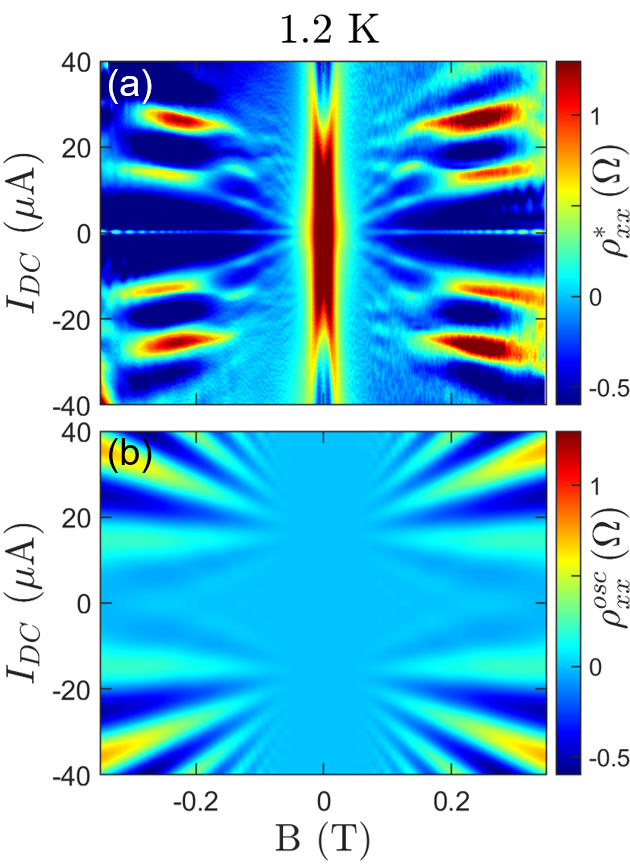}
\caption{(a) Plot of $\rho^{*}_{xx}$ data for 1.2 K, and (b) corresponding plot of oscillatory component of PIROs $\rho^{osc}_{xx}$ calculated from model in \cite{dmitriev2010}. See discussion in main text regarding Fig. \ref{fig:bigfig}.}
\label{fig:bigfigextra} \end{figure}

\textit{Appendix C: 1.2 K data}---Figure \ref{fig:bigfigextra} shows (a) data at the intermediate temperature of 1.2 K and (b) matching calculation highlighting the transition in the subsonic regime from HIRO-dominated at 10 mK to PIRO-dominated at 3.9 K (see Fig. \ref{fig:bigfig}).

\textit{Appendix D: Signs of $M$, $p$, $\epsilon_{DC}$ and $\epsilon_{ph}$}---We comment on the signs of the HIñRO order $M$, PIRO index $p$, and the variables $\epsilon_{DC}$ and $\epsilon_{ph}$ following the definitions given in the text. $M$ and $\epsilon_{DC}$ are positive (negative) for the $+B$ and $+I_{DC}$, and the $-B$ and $-I_{DC}$ ($-B$ and $+I_{DC}$, and the $+B$ and $-I_{DC}$) quadrants. $p$ and $\epsilon_{ph}$ are positive (negative) for $+B$ ($-B$). Integer-valued $M$ and $p$ respectively represent special values of the continuous variables $\epsilon_{DC}$ and $\epsilon_{ph}$. We draw attention that the transformation from the $(B,I_{DC})$ coordinate system to the $(\epsilon_{ph},\epsilon_{DC})$ coordinate system straightforwardly maps resistivity data from the $+B$ and $+I_{DC}$, and the $+B$ and $-I_{DC}$ quadrants to the $\emph{same}$ quadrants in $(\epsilon_{ph},\epsilon_{DC})$ space. However, resistivity data from the $-B$ and $+I_{DC}$, and the $-B$ and $-I_{DC}$ quadrants are $\emph{inverted}$ in $(\epsilon_{ph},\epsilon_{DC})$ space, i.e., data in the $-B$ and $+I_{DC}$, and the $-B$ and $-I_{DC}$ quadrants respectively are mapped to the $-\epsilon_{ph}$ and $-\epsilon_{DC}$, and the $-\epsilon_{ph}$ and $+\epsilon_{DC}$ quadrants.

\textit{Appendix E: Summary of scattering processes depicted in Fig. \ref{fig:summary}(e)}---The Hall electric field causes the LLs to become tilted in real space along the y-direction, i.e., the LL energies vary linearly along the direction of the electric field. In this tilted LL landscape, resonant scattering events by phonons can in general take electrons to higher or lower energy accompanied by a change in the guiding center position. Crucially, the allowed resonant transitions have different character in the three regimes of interest. Defining ``downstream'' and ``upstream'' as the direction of the scattering event relative to the tilt, the three regimes are as follows. Subsonic regime: The phonons mediate four possible inter-LL scattering processes (absorption/emission, upstream/downstream), but all four processes are suppressed as the temperature is reduced. Critical point at ``sound barrier'': The tilt of the LLs matches the slope of the phonon resonance condition for the downstream emission and upstream absorption processes which no longer connect different-index LLs- instead \emph{intra}-LL scattering is possible. There is now no hinderance to the downstream emission process- the top most LL near the Fermi level will always have some vacancies. Supersonic regime: When the electrons drift faster than the emitted phonons can propagate, the downstream emission process (from a lower occupied to higher unoccupied index LL) dominates and is not surpressed as the temperature is reduced. See SM \cite{supplemental} Sec. S3 for full discussion. 

\textit{Appendix F: Electron heating}---Electron heating is an important factor in our experiments. Even for \emph{zero DC current}, at the 10 mK-base temperature of the dilution refrigerator, from separate measurements under equilibrium conditions, the (effective) electron temperature is estimated to be $T_0\simeq$40 mK \cite{paper1}. Electron heating is inevitably enhanced further in high current measurements. However, crucially we can exclude Joule heating, since the current power injected into the Hall bar is much smaller than the cooling power of the mixing chamber, even at base temperature- see SM \cite{supplemental} Sec. S1. In our earlier work on nonlinear transport phenomena, also in an ultrahigh mobility 2DEG Hall bar with a similar geometry, we determined an empirical relation for the electron temperature $T_e$ dependence as a function of $I_{DC}$ in \textmu A of $T_e\simeq T_{0}+1.0\, I_{DC}^{0.65}$ \cite{paper1}. With that relation, for a 15 \textmu A current (corresponding to the ``sound barrier'') we estimate an electron temperature of $\sim$6 K (and for 40 \textmu A $\sim$11 K) for the 10 mK data. Yet, this elevated electron temperature does not imply an equal increase in the phonon temperature and we expect that the electron temperature can differ substantially from the phonon temperature. Nonetheless, the higher electron temperature is likley a factor in the observed fall-off in the PIRO amplitude as $I_{DC}$ approaches 40 \textmu A- see also SM \cite{supplemental} Fig. S3 and Figs. S5 (b) and (c). This fall-off may also be a consequence of a reduction of the scattering times due to a higher electron temperature- see SM \cite{supplemental} Sec. S4. A theory incorporating electron heating whereby heat trasfer between the electrons and phonons is included would be beneficial. However, we stress that current theories, such as the one by Dmitriev et al. \cite{dmitriev2010}, only use \emph{one} temperature to describe both the phonons and electrons. The full impact of electron heating especially deep in the supersonic regime is left for future work.

\include{allsuppl.tex}

\end{document}

%% file: biblio.tex

\bibliographystyle{apsrev4-2}
\bibliography{refs}

%% file: allsuppl.tex
\onecolumngrid
\vspace{1\baselineskip}
\begin{center}
  {\Large \bfseries Supplemental Material: \\ Resonant Magneto-phonon Emission by Supersonic Electrons in Ultra-high Mobility Two-dimensional System}
\end{center}
\twocolumngrid

\renewcommand{\thesection}{S\arabic{section}} 
\renewcommand{\thefigure}{S\arabic{figure}}

\section{\uppercase{Background subtraction}}  \label{sec:bg}

\begin{figure*}[]
\includegraphics[width=2\columnwidth]{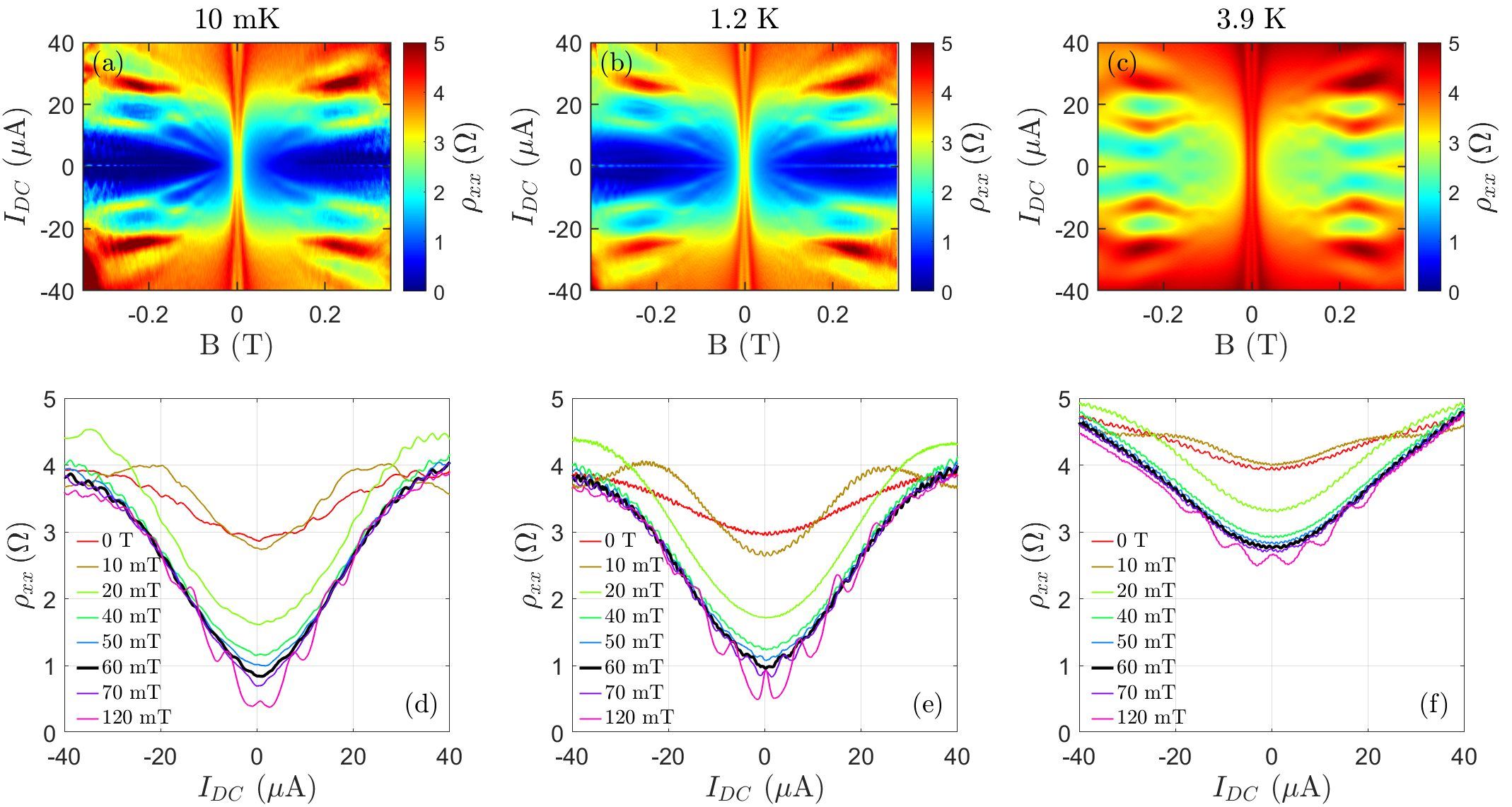}
\caption{Top row: Raw differential resistivity $\rho_{xx}(B, I_{DC}$) plots at (a) 10 mK, (b) 1.2 K and (c) 3.9 K. Bottom row: $\rho_{xx}$ versus $I_{DC}$ traces for selected $B$-fields up to 120 mT at (d) 10 mK, (e) 1.2 K and (f) 3.9 K. For each of the three data sets, we take the optimal field that reflects (just) the background increase best to be 60 mT, and henceforth subtract the (bold black) 60 mT trace, \emph{after it is smoothed}, from $\rho_{xx}(B, I_{DC}$) to obtain $\rho_{xx}^{*}(B, I_{DC})$.}
\label{fig:bg}
\end{figure*}

For the data presented, to focus on the resonant features of interest, we found it convenient to examine the differential resistivity $\rho_{xx}^{*}$, which we define as the raw differential resistivity $\rho_{xx}(B, I_{DC})$ minus, at each temperature, a smooth background $\rho_{xx}^{bg}$ that is approximately parabolic with respect to DC current for currents up to $\sim$ 20 \textmu A. In Fig. \ref{fig:bg}, we show the raw data plots for $\rho_{xx}$ at (a) 10 mK, (b) 1.2 K, and (c) 3.9 K. The general increase in $\rho_{xx}$ with respect to the DC current is present at all three temperatures over the full $B$-field range examined other than very close to zero field. The background increase, when approximately parabolic, can be attributed to enhanced electron-electron scattering with increasing DC current- see Ref. \cite{paper1} for further discussion. This effect was observed in our previous measurements on a Hall bar fabricated from a different GaAs/AlGaAs hetero-structure on application of a DC current up to 10 \textmu A \cite{paper1}. In Fig. \ref{fig:bg}, we show selected $\rho_{xx}$ versus $I_{DC}$ traces for certain values of $B$ at (d) 10 mK, (e) 1.2 K and (f) 3.9 K. Within a couple tens of milli-Tesla of zero field the traces are strongly influenced by a double-peak feature located on top of the large nMR and the nMR itself centered at zero field \cite{paper1}. Thereafter, for fields just above approximately 40 mT the smooth background increase in $\rho_{xx}$ is clear (unfettered by other features). The smooth background increase with current has barely any $B$-field dependence at this point. At sufficiently high $B$-field (clear in the 120 mT traces), the background increase becomes obscured as SdH oscillations at zero DC current, and HIROs at 10 mK and PIROs at 3.9 K at finite DC current, develop. Consequently, for each of the three temperatures, $\rho_{xx}^{bg}$ is taken to be the \emph{smoothed} $\rho_{xx}(B=60$ mT$, I_{DC})$ trace. The choice of the 60 mT trace ensures that the subtraction procedure for the background does not incorporate artifacts of the nMR at 0 T, nor the SdH oscillations, HIROs, or PIROs that develop at higher $B$-field- see bold black 60 mT traces in Fig. \ref{fig:bg} panels (d)-(f). Note that 60 mT is also taken as the threshold for the onset of PIROs relevant to the fit discussed in Sec. \ref{sec:gsquared}. We stress that the background removal procedure enables us to better compare features in $\rho_{xx}^{*}$ that originate from resonant scattering processes involving acoustic phonons with features in the calculated electron-phonon transport scattering rate following the theory in Ref. \cite{dmitriev2010}.

In our experiments here, Joule heating due to the applied DC current is minimal since the estimated maximum DC power dissipated is less than 1.6 $\mu$W which is small compared to the cooling power of the dilution refrigerator at base temperature. Also, we do not observe any heating in the thermometer attached to the mixing chamber plate at small magnetic fields ($B \le$ 0.35 T) where the two-terminal resistance remains less than 1 k$\Omega$.

\section{\uppercase{Hall Induced Resistance Oscillations}}  \label{sec:hiro}

\begin{figure}[]
\includegraphics[width=0.9\columnwidth]{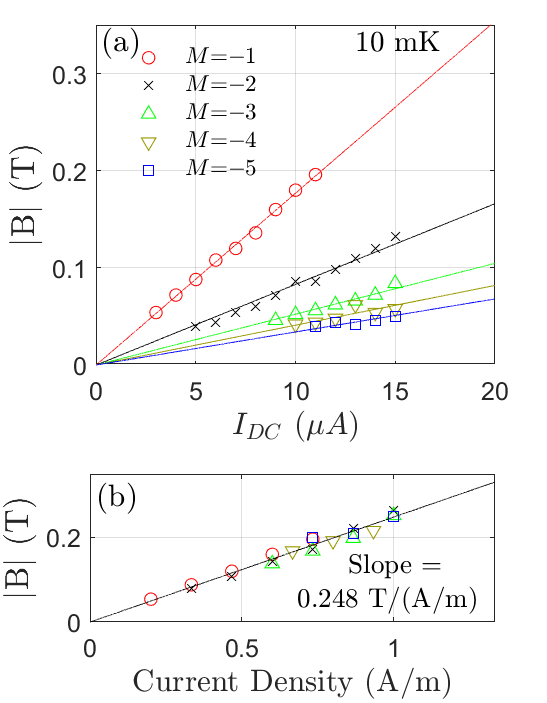}
\caption{(a) Position of HIRO maxima in $\rho_{xx}^{*}$ for HIRO peaks up to fifth order ($M$=$-5$). Points are taken from the 10 mK data set in Fig. 1(a) for the $+I_{DC}$ and $-B$ quadrant. (b) The HIRO maxima position fall on a common line obeying the relation $B\cdot M \propto I_{DC}/W$ namely the current density. Taking $W$=15 $\mu$m, from the slope of the common line we determine the parameter $\gamma$=$1.95\pm0.04$.}
\label{fig:gamma}
\end{figure}

In the 10 mK data in Fig. 1(a), HIROs become detectable around $B=0.06$ T, and are characterized by resonances that fan out from the origin. HIROs have been explained through a Zener-like tunneling process for electrons hopping between Landau levels that are spatially tilted by the Hall electric field (along the y-direction perpendicular to the applied DC current) \cite{yang2002zener,Bykov2005,zhang2007MT,zhang2007effect,vavilov2007,Lei2007,bykov2008,Hatke2009,vitkalov2009NL,Hatke2010,Hatke2011,Hatke2012,Shi2014b,hatke2015SAND,shi2017,Zudov2017,greenaway2021,paper1,Bartels2025}- see black arrows in Fig. 1(e) cartoons. Here we verify that the HIROs observed in the subsonic regime are well behaved following an established model, and demonstrate that the approximate expression we use for $\epsilon_{DC}$ is valid. Following the theory in Ref. \cite{yang2002zener}, HIROs are expected to obey the formula $\epsilon_{DC} = \gamma I_{DC}k_F/enW\omega_c =M=$ $\pm$1, $\pm$2, $\pm$3, $\ldots$. In this formalism, $\gamma$ takes a value of approximately 2 \cite{Bartels2025}. Examining carefully the position of the extracted HIROs maxima in the 10 mK data, we observe HIROs maxima up to fifth order, find the HIROs collapse on to a common line when plotting their position in $B$-field versus current density, and determine $\gamma = 1.95 \pm 0.04$- see Fig. \ref{fig:gamma}. Consequently, following the assumption made about the value of $\gamma$ in Ref. \cite{dmitriev2010}, we take $\gamma = 2$ when we calculate the $\epsilon_{DC}$ values for the $\epsilon_{DC}$ versus $\epsilon_{ph}$ plots in Fig. 3. Notice that at 10 mK the HIROs appear in Fig. 3 as horizontal line features at nearly integer values of $\epsilon_{DC} =$ $\pm$1, $\pm$2, $\pm$3, $\ldots$. A $\gamma = 2$ value for the quasi-elastic inter-Landau-level scattering process responsible for HIROs is equivalent to picturing the dominant scattering to be with wave-vector change of 2$k_F$ (or equivalently the electron's guiding center is displaced by a distance 2$R_c$ along the y-direction)- see Fig. 1(e) \cite{Bartels2025}. Similarly, also note that inter-Landau level scattering by emission or absorption of phonons for PIROs is commonly pictured to occur with exact wave-vector change of 2$k_F$ which is also an approximation \cite{zhang2008,dmitriev2010}.

The black HIRO lines in panels (c) and (d) of Fig. 1 are calculated from the expressions $\epsilon_{DC} = \gamma I_{DC}k_F/enW\omega_c$ and $\omega_c= eB/m^{*}$ \cite{yang2002zener}, taking parameters $\gamma$=2, $W=$ 15 \textmu m, $n$ = 2.2 $\times$ 10$^{11}$ cm$^{-2}$, and $m^{*}=0.067m_{e}$ for the effective mass where $m_{e}$ is the mass of an electron.

Previously, it has been reported that HIROs decay not only with increasing temperature \cite{zhang2008,Hatke2009}, but also with increasing DC current (at fixed $B$-field) \cite{yang2002zener,zhang2007effect,Lei2007,hatke2015SAND,paper1,Bartels2025}- see also the theoretical discussion in Ref. \cite{Lei2007}. As expanded on below in Sec. \ref{sec:acoustic}, Hatke \textit{et al.} observed in the differential resistivity a gradual suppression of HIROs on approaching the ``sound barrier'', and a broad peak at the ``sound barrier'' at low field (and down to zero field) with no HIRO features evident beyond this broad peak \cite{hatke2015SAND}. For the 10 mK data shown in Fig. 1(a) and Fig. 3(ai), the HIRO lines likewise appear to terminate abruptly at the ``sound barrier'' and are absent in the supersonic regime (where PIRO-behavior is dominant).

\section{\uppercase{RESONANT SCATTERING PROCESSES INVOLVING ACOUSTIC PHONONS}}  \label{sec:acoustic}

\begin{figure}[ht]
\includegraphics[width=1\columnwidth]{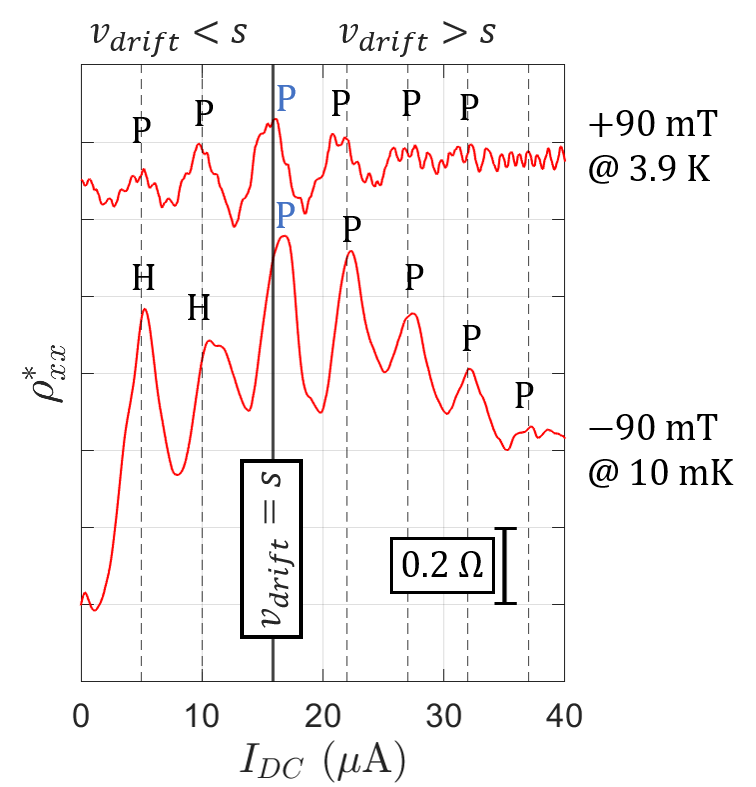}
\caption{$\rho_{xx}^{*}$ versus $I_{DC}$ trace at -90 mT (+ 90 mT) from the 10 mK (3.9 K) data. Traces are vertically offset for clarity. P (H) marks a PIRO (HIRO) peak. Refer to panels (a)-(d) in Fig. 1. The solid vertical line identifies the ``sound barrier'', and the approximately equally spaced dashed vertical lines act as a guide to the eye. In the subsonic regime the peaks are HIROs (PIROs) in origin at 10 mK (3.9 K), whereas in the supersonic regime they are PIROs in origin at both temperatures. Note that for the choice of $B$-field here, HIROs and PIROs are approximately equally spaced. We strongly stress that in distinction to a single broad peak observed at the ``sound barrier'' in the work of Hatke \textit{et al.} \cite{hatke2015SAND}, we have observed PIRO oscillations in the supersonic regime. To emphasize the special status of the peak at the ``sound barrier'', namely the peak is treated as the zero-order PIRO peak (with index $p=$ 0) within a cyclotron-orbit picture, we have colored the P label blue.  We also comment that in the 3.9 K trace, the PIROs amplitude is strongest at the ``sound barrier'' and diminishes to both lower and higher DC current. This characteristic is consistent with the calculations in Ref. \cite{dmitriev2010}. Lastly, strongly note that counter to the theory in Ref. \cite{dmitriev2010} which predicts that the amplitude of PIROs saturates at very low temperature, we observe approximately a factor of two enhancement in the amplitude of PIROs in the 10 mK data compared to the 3.9 K data.}  

\label{fig:traces90mT}
\end{figure}

Since for the experimental conditions discussed the maximum cyclotron energy is $\sim$0.6 meV, we are solely concerned with acoustic phonons and can neglect optical phonons which have much a much higher energy of $\sim$37 meV. The general properties of the four \emph{resonant} scattering processes involving acoustic phonons, as represented by the four red arrows in the cartoons in Fig. 1(e), are as follows. Two of the processes that promote electrons to higher energy involve phonon absorption and two of the processes that demote electrons to a lower energy involve phonon emission. At zero DC current (LLs not tilted), the two phonon absorption (emission) processes \emph{equally} take electrons from the $N$ th LL to the $N+p$ th ($N-p$ th) LL. At finite DC current (LLs tilted) the four scattering processes are additionally distinguished by whether the scattering of the electrons, equivalently the displacement of the guiding center, is in the direction along (``downstream'') or against (``upstream'') the tilted LLs. At say a fixed given $B$-field, for increasingly tilted LLs, the downstream absorption and the upstream emission processes (pair I) progressively can span ever more LLs \emph{always} taking electrons to a higher index LL and lower index LL respectively. The situation for the downstream emission and the upstream absorption processes (pair II) is very different. For increasingly tilted LLs, in the subsonic regime, these two processes will span fewer LLs on taking electrons to a higher index LL or a lower index LL. At some point, at the ``sound barrier'', when the LLs become parallel to the pair-II-process arrows, only intra-LL transitions are possible. However, on entering the supersonic regime at higher DC current, inter-LL transitions are possible again but counterintuitively, now with highly tilted LLs, the downstream emission process can take electrons from the (mostly occupied) $N$ th LL to an unoccupied higher index LL. This process is responsible for the strong PIRO signal in the supersonic regime that is only weakly sensitive to temperature at mK temperatures \cite{dmitriev2010}. Note that in contrast, the upstream absorption process is even more strongly suppressed in the supersonic regime at very low temperature. Not only are there hardly any phonons to absorb from the lattice bath but electrons in the $N$ th LL will have extremely low probability to find unoccupied states in a lower index LL (that is well below the Fermi level).

For positive and negative integer values of $p$, the loci of the red PIRO lines in panels (c) and (d) of Fig. 1 are calculated from the expression $pB_{p} = 2\left(\frac{nhs}{ev_f} \pm \frac{h}{e^2}\frac{I_{DC}}{v_FW}\right)$ with $s=$ 3.0 km/s ($v_{drift}$ at the ``sound barrier'' equivalent to $I_{DC} = 15.9$ $\mu$A: see Fig. 4 and discussion). Here $B_{p}$ is the $B$-field position of the PIRO line with index $p$, $h$ is Planck's constant and $v_F$=$\hbar k_F / m^{*}$ is the Fermi velocity. Each red line represents a pair of allowed scattering processes (pair I or pair II as discussed above). At zero DC current, lines with the same index $p$ (with positive and negative sign) converge to points (at positive and negative $B$) giving the location of conventional PIRO peaks. At finite DC current, lines with different $p$-index also can cross. In either case, a crossing of two red PIRO lines identifies a point where all four of the possible resonant scattering processes can be active leading to an enhanced peak in $\rho_{xx}^{*}$. In either the positive or negative $B$-field sector, suppose we apply a positive current, then a red line moving to higher absolute $B$-field (towards the common point at zero field here at $I_{DC}$ = +15.9 $\mu$A) tracks the pair-I (pair-II) processes. We stress that the interesting phonon physics occurs when a pair-II line in the subsonic regime for one $B$-field polarity passes through the common point at zero field to the opposite $B$-field polarity and enters the supersonic regime. In the subsonic regime, the downstream emission process is strongly suppressed at low temperature because the lower-index LLs into which electrons must scatter are almost fully occupied. In contrast, in the supersonic regime, this process proceeds easily by virtue of the strongly tilted LLs because electrons can now easily scatter into virtually empty higher-index LLs. The same arguments apply for negative current noting that inverting the current changes the LL tilt direction and so a line representing pair-I-type processes for say positive current polarity will now represent pair-II-type processes for negative current polarity. We refer the interested reader to the insightful discussion in Ref. \cite{greenaway2021} on the splitting of the magneto-phonon resonances observed in a graphene Hall bar at zero DC current on application of a finite DC current and the induced shifting with analogy to the Doppler effect. We strongly emphasize that the calculated red PIRO lines should be treated as guide lines. Within a simple picture they merely map out in $B$ versus $I_{DC}$ (equivalently $\epsilon_{ph}$ versus $\epsilon_{DC}$) space energetically allowed resonant transitions between tilted and un-tilted Landau levels involving acoustic phonons with velocity $s$ assuming $2k_F$ momentum transfer. Whether a PIRO peak is observable or not at or near an expected resonant condition depends on the ambient phonon population in the lattice bath, the availability of empty electronic states near the Fermi energy, and the competition between scattering channels involving phonons and those involving other mechanisms (such as the quasi-elastic scattering channels responsible for HIROs). Furthermore, as highlighted in Sec. \ref{sec:phasechange}, there are subtle effects related to phase that can shift a PIRO feature away from its expected position- see Refs. \cite{dmitriev2010,hatke2011piro}.     

The value of $s=3.0$ km/s that we have determined from our data is consistent with values in the range of 2.9-5.9 km/s reported in the literature for PIROs in GaAs-based hetero-structures \cite{zudov2001,yang2002A,zhang2008,bykov2005MP,hatke2009phonon,Bykov2009,bykov2010,hatke2011piro,hatke2015SAND,raichev2017}. The value can depend on the type of dominant phonon modes present (their character and propagation direction with respect to the crystallographic axes, and whether they are interface-like or bulk-like \cite{Ponomarev2001,zudov2001,Ryzhii2003,Zhang2004XYZ,Lei2008}) and the exact composition of the hetero-structure. Our data is well explained by a single phonon mode rather than two (or more) modes likely because of the ultra-high mobility of the 2DEG employed and the low temperature range (10 mK- 3.9 K) studied \cite{zudov2001,raichev2009,hatke2015SAND,kumara2019,Greenaway2019}.

Hatke \textit{et al.} have reported for a lower mobility 2DEG a broad peak in the differential resistivity \emph{at zero magnetic field} ($\epsilon_{ph} \rightarrow \pm \infty$) at the ``sound barrier'' for 25 and 50 $\mu$m wide Hall bars \cite{hatke2015SAND}. The peak was subsequently explained in terms of a Bloch-Gr{\"u}neisen non-linearity arising from change to the electron-phonon scattering rate by current-induced heating \cite{raichev2017}, i.e., by a mechanism different from PIROs. In our case, for a narrower Hall bar, the prominent nMR and the double-peak feature centered at zero $B$-field obscure clear observation of such an effect [see red colored $\rho_{xx}$ traces in panels (d)-(f) in Fig. \ref{fig:bg} which nonetheless are quite non-linear with DC current]. We stress that our study addresses aspects not reported or addressed in Refs. \cite{hatke2015SAND,raichev2017}, namely: i. the interplay between PIRO and HIRO \emph{throughout} the subsonic regime as the temperature is changed; ii the clear observation of periodic oscillations due to PIROs in the supersonic regime at finite magnetic field (see Fig. 1 and Fig. \ref{fig:traces90mT}); and iii. a detailed comparison of experimental data with the theory of Dmitriev \textit{et al.} (including evidence for the predicted $\pi/2$ phase change effect on transitioning from the subsonic to the supersonic regime discussed in Sec. \ref{sec:phasechange}) \cite{dmitriev2010}.  

On the one hand, the standard picture for PIROs and the model of Dmitriev \textit{et al.} in Ref. \cite{dmitriev2010} address implicitly $\emph{resonant}$ (2$k_F$) phonon scattering at finite magnetic field, i.e., in a cyclotron-orbit picture, Landau levels are an essential ingredient to support inter-Landau level transitions. Determined by the value of the quantum lifetime $\tau_q$ (equivalently $B_q$), PIROs are expected to wash out as the $B$-field approaches zero when Landau levels are no longer distinct- see Figure 2 both top and bottom rows. On the other hand, the explanation for the robust broad peak observed by Hatke \textit{et al.} over a 2-12 K range at the ``sound barrier'' at zero field \cite{hatke2015SAND} evidently does not require a magnetic field \cite{raichev2017}, i.e., a cyclotron-orbit (or Landau-level-quantization) picture is not essential and the peak can in principle occur at any magnetic field at a DC current independent of $B$ set only by the condition $v_{drift}=s$. Indeed, Ref. \cite{hatke2015SAND} reports the presence of the broad peak with field-independent position at $v_{drift}=s$ from zero field up to at least 125 mT (but no other peaks are seen at higher DC current). In contrast, we observe (from $\sim$60 mT) up to higher $B$-field, a strong resonance at $v_{drift} \sim s$, and several other peaks at higher DC current (Fig. \ref{fig:traces90mT}) that weaken as the $B$-field approaches zero (and become obscured by other features near zero field such as the nMR). Although unquestionably acoustic phonon-related, our observations and those in Ref. \cite{hatke2015SAND} at the ``speed of sound'', relate to two subtlety different mechanisms at opposite ends of a spectrum. We are not aware of a unified model that marries the theoretical descriptions in Ref. \cite{dmitriev2010} (cyclotron-orbit picture) and Ref. \cite{raichev2017} (zero-field picture). In the main text, as it is more appropriate for our data and observations, we have essentially invoked the cyclotron-orbit picture, and treated the ``sound barrier'' features as a special case of PIROs, namely, the thick horizontal red lines in panels (c) and (d) of Fig. 1 are generated from the above mentioned expression for $B_{p} (I_{DC})$ by assigning an index $p=$ 0 (note no $p=$ 0 index PIRO peak occurs at zero DC current), and, as conveyed by the magenta boxed cartoon in Fig. 1(e), resonant phonon scattering responsible for ``sound barrier'' features is pictured as \emph{intra}-Landau level scattering by emission of phonons with any wave-vector up to $\sim$2$k_F$ (depicted by short thick red arrows). Finally, in the PIRO picture, as the $B$-field tends to zero, all PIRO lines (including the two with index $p=$ 0) converge to one of two special points at zero field with $I_{DC}$ value set by the condition $v_{drift} =\pm s$ consistent with the zero-field picture.

\section{\uppercase{FIT OF SUBSONIC 3.9 K DATA AND DISCUSSION OF PARAMETERS}}  \label{sec:gsquared}

\begin{figure}[]
\includegraphics[width=0.9\columnwidth]{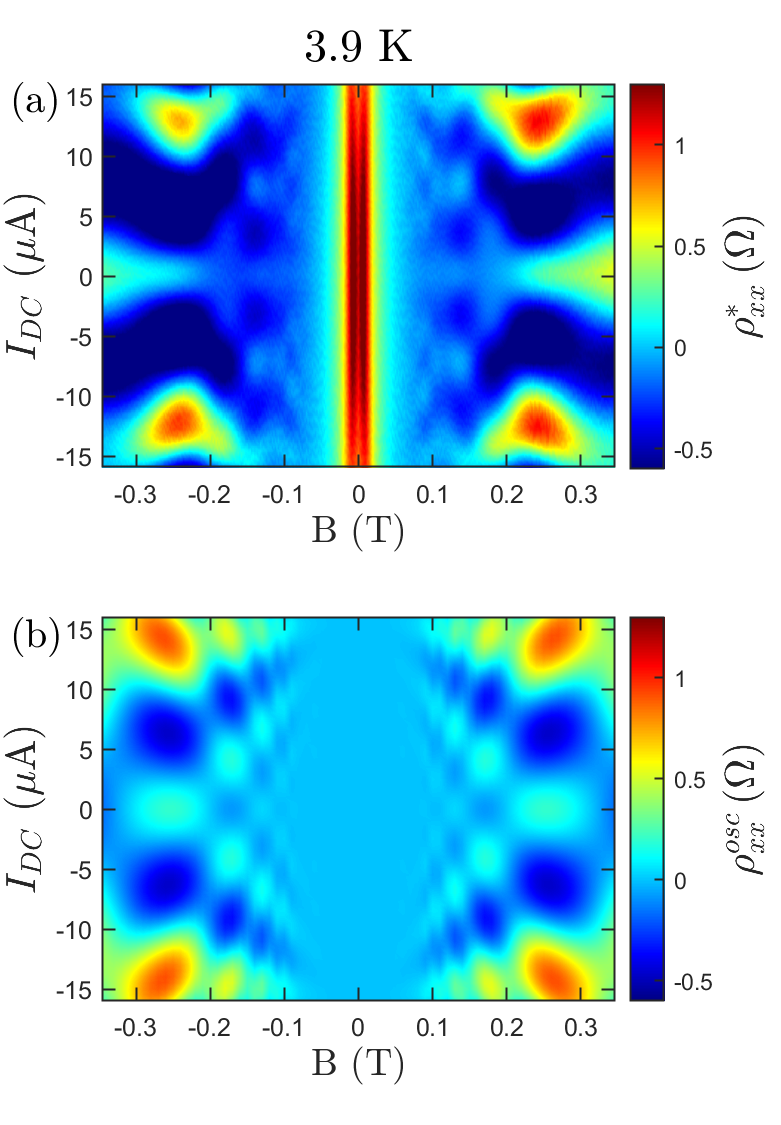}
\caption{(a) $\rho_{xx}^{*}$ data at 3.9 K in the subsonic regime, and (b) theory using Eq. (5) from the model in Ref. \cite{dmitriev2010}. See text for details of, and rational for, the fit in the subsonic regime, along with the parameters.}  

\label{fig:pirosubsonic}
\end{figure}

We outline here our approach to calculating the $\rho^{osc}_{xx}$ plots on the bottom row of Fig. 2 and Fig. 5. In order to apply the theory of Ref. \cite{dmitriev2010}, we need two key parameters: the quantum lifetime $\tau_q$ and the electron-phonon coupling constant $g^2$. The former we can first estimate independently from the onset of PIROs, whereas the latter we can estimate from a best fit of the 3.9 K $\rho_{xx}^{*}$ data restricted to the subsonic regime dominated by PIROs assuming an appropriate value for $\tau_q$. The subsonic regime data at 3.9 K is used to determine $g^2$ because the PIRO resonance features are most pronounced at this temperature, where the impact of HIROs and SdH oscillations are minimal, and the influence of current-induced electron heating is less significant (see Appendix F in the main text). Since the processes responsible for the nMR and the double-peak feature centered at zero field are not incorporated in the theory in Ref. \cite{dmitriev2010}, we have excluded data in the range of -60 mT to +60 mT from the fit. Furthermore, we have fitted the data just in the subsonic regime where the behavior of PIROs is better established and documented \cite{zhang2008,greenaway2021}, and where a comparison is easier to make. Since it has yet to be tested fully against experimental data, the model in Ref. \cite{dmitriev2010} is likely subject to more uncertainty far from equilibrium at higher DC current in the supersonic regime. As a last step, we take the determined values of $\tau_q$ and $g^2$, assume they vary little with temperature and do not depend appreciably on the value of the DC current, and calculate the $\rho^{osc}_{xx}$ plots that extend into the supersonic regime up to 40 $\mu$A. Other input parameters for the fit of the 3.9 K data are: $m^{*}=0.067m_{e}$ for the effective mass, the Hall bar width $W=$ 15 $\mu$m, the 2DEG carrier density $n$ = 2.2 $\times$ 10$^{11}$ cm$^{-2}$, the mobility $\mu$=18 $\times$ 10$^6$ cm$^2$/(Vs), the sound velocity $s=$ 3.0 km/s, and the temperature $T=$ 3.9 K.

A value of $\tau_q=$ 6.3 ps is determined from the onset of PIROs at $B_q =$ 0.06 T in the 3.9 K data [see Fig. 1(b) and Fig. \ref{fig:bg}(f)] following the expression $\omega_c\tau_q=1$. For the 2DEG studied, the onset of HIROs and the onset of PIROs are very similar- see respectively panels (d) and (e) in Fig. \ref{fig:bg} where the onset is evident on close inspection of the 60 and 70 mT traces. Following Ref. \cite{dmitriev2010} and using their Eq. (5), we determine $g^2$ directly by fitting the 3.9 K $\rho_{xx}^{*}$ data in the subsonic regime (see Fig. \ref{fig:pirosubsonic}) and obtain a best fit value of $g^{2}=$ 0.0016 $\pm$ 0.0002. 

We can now use this electron-phonon coupling constant $g^{2}$ to estimate the electron-phonon scattering rate and make a comparision to other reported values of the electron-phonon scattering rate. At a sufficiently high temperature, Dimitriev \textit{et al.} \cite{dmitriev2010} relate $g^2$ to the the non-oscillating (n-o) part of the electron-phonon scattering time using $\tau_{e-ph}^{(n-o)} = \hbar/g^2k_BT$ where $k_B$ is the Boltzmann constant. Using our best fit value of $g^{2}$=0.0016 leads to an expected $\tau^{(n-o)}_{e-ph} = 4.8$ ns/$T$[K], which can be compared to the zero DC current transport ($tr$) electron-phonon scattering time $\tau_{e-ph}^{(tr)}$ from the linear temperature dependence typically observed above 1 K in high mobility 2DEGs. For instance, for 2DEGs with similar densities to the 2DEG we study, in Ref. \cite{Keser2021}, $\rho\sim \alpha T$ with $\alpha \sim 0.84$ $\Omega$/K (for $n = $ 1.8 $\times$ 10$^{11}$ cm$^{-2}$) was found above 1 K (but below the Bloch-Gr{\"u}neisen temperature). See also Ref. \cite{Ashlea2024} and Ref. \cite{shi2014} respectively where linear dependencies with $\alpha \sim 0.64$ $\Omega$/K (for $n = $ 2.45 $\times$ 10$^{11}$ cm$^{-2}$) and $\alpha \sim 0.91$ $\Omega$/K (for $n = $ 2.0 $\times$ 10$^{11}$ cm$^{-2}$) are reported. In our Hall bar device, above 1 K, we find a very similar linear temperature dependence for $\rho$ with $\alpha \sim 0.75$ $\Omega$/K estimated for $B \geq$ 0.1 T (away from the nMR peak) and at zero DC current (noting that the Bloch-Gr{\"u}neisen temperature is $\sim$5.1 K \cite{raichev2017}). Assuming that the main contribution is due to $tr$ electron-phonon scattering, we extract $\tau_{e-ph}^{(tr)} =$ 1.6 ns/$T$[K] in this regime using the Matthiessen rule $\rho_{xx}$ = $(m^{*}/ne^2)(1/\tau_0+1/\tau_{e-ph}^{(tr)})$, where 1/$\tau_{e-ph}^{(tr)}$ is linear in temperature and $\tau_0$ is a temperature independent constant. Compared to $\tau^{(n-o)}_{e-ph}$, $\tau_{e-ph}^{(tr)}$ is about three times smaller in value. We stress that the quantity $\tau^{(n-o)}_{e-ph}$ is determined in the strong non-linear transport regime featuring resonant phonon emission and absorption, whereas the quantity $\tau_{e-ph}^{(tr)}$ is determined in the linear transport regime. Consequently different processes contribute to the determination of the two quantities which can explain the difference in the values.

\section{\uppercase{phase change effect transitioning from subsonic regime to supersonic regime}}  \label{sec:phasechange}

\begin{figure*}[]
\includegraphics[width=2\columnwidth]{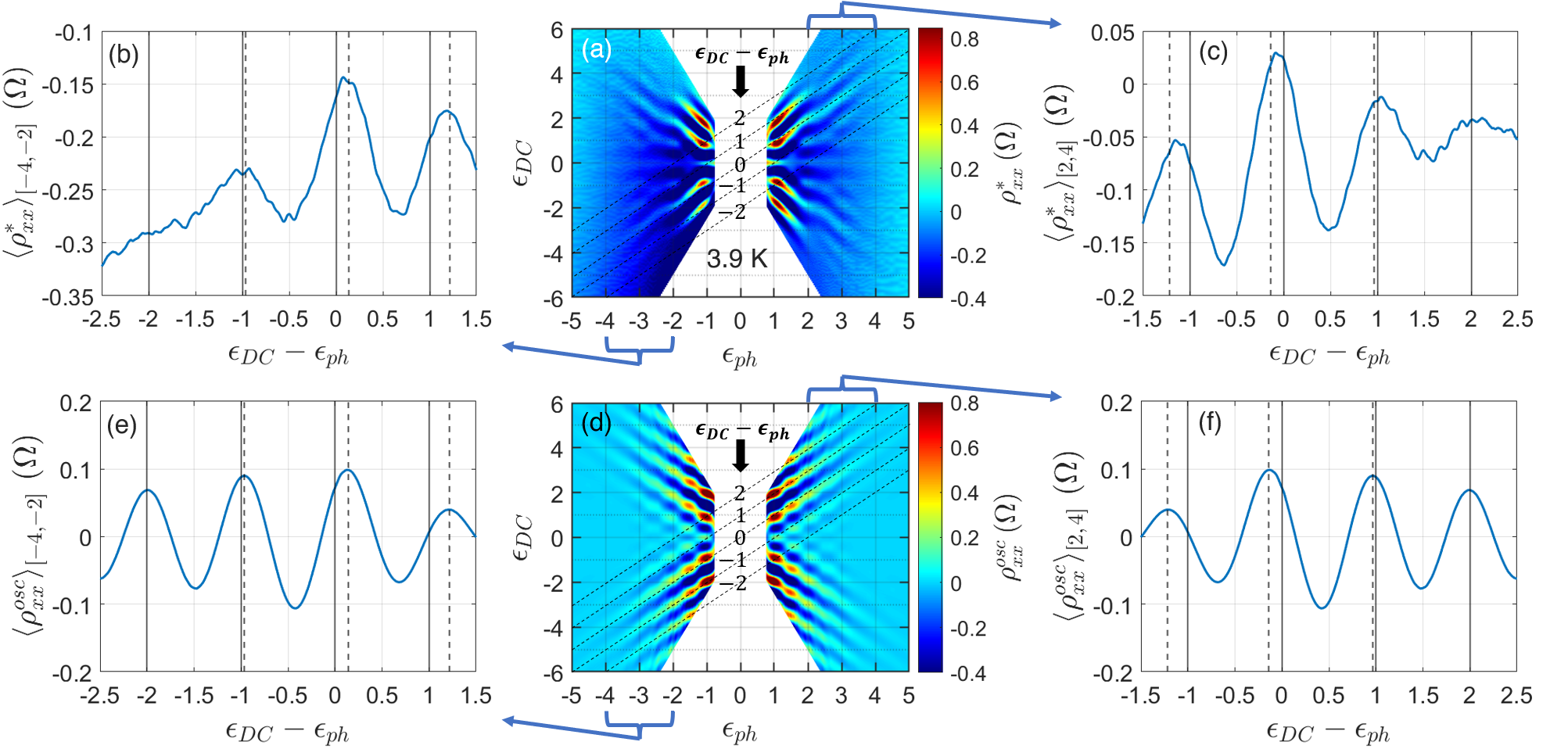}
\caption{Evidence for phase change effect in PIROs across the ``sound barrier'' (at $v_{drift} = \pm s$ or equivalently $\epsilon_{DC} \pm \epsilon_{ph} = 0$). Top row: (a) $\rho_{xx}^{*}$ from the 3.9 K experimental data replotted in $\epsilon_{ph}$ versus $\epsilon_{DC}$ space. Diagonal black dashed lines identify integer values of $\epsilon_{DC}-\epsilon_{ph}$ (= -2, -1, 0, +1, +2) for which PIRO oscillation maxima are anticipated to occur according to a simple picture of PIROs. Note that for clarity we have omitted the corresponding anti-diagonal lines. (b) and (c) show respectively traces of $\rho_{xx}^{*}$ averaged over $\epsilon_{ph} \in [-4, -2]$ and $\epsilon_{ph} \in [+2, +4]$ versus $\epsilon_{DC}-\epsilon_{ph}$- see text for further discussion. Bottom row: (d) $\rho_{xx}^{osc}$ calculated following the model in Ref. \cite{dmitriev2010} in $\epsilon_{ph}$ versus $\epsilon_{DC}$ space. The calculation assumes the same parameters taken for the plots shown in Fig. 2(d) and Fig. 3(bii). Diagonal black dashed lines identifying integer values of $\epsilon_{DC}-\epsilon_{ph} = -2, -1, 0, +1, +2$ are also added. (e) and (f) show respectively traces of $\rho_{xx}^{osc}$ averaged over $\epsilon_{ph} \in [-4, -2]$ and $\epsilon_{ph} \in [+2, +4]$ versus $\epsilon_{DC}-\epsilon_{ph}$. Vertical dashed lines in plots (b), (c), (e) and (f) identify the position of the calculated PIRO maxima to illustrate the predicted phase change effect. Note that panel (c) of Fig. 1 in the main text shows the two traces presented in panels (c) and (f) from this figure here.}
\label{fig:phase}
\end{figure*}

Experimental studies of PIROs most conveniently capture magneto-resistance data as a function of $B$ and $I_{DC}$, either by sweeping $I_{DC}$ and stepping $B$, as in our case, or vice versa. At sufficiently high temperature such that PIROs dominate [see 3.9 K data in Fig. 1(b)], in both the subsonic and supersonic regimes, PIRO features, in a simple picture, are expected to essentially track the red solid lines in Fig. 1(d), albeit with amplitude enhanced in the vicinity where two of the red lines intersect at zero or finite DC current. As noted above, the red lines of differing slope converge to one of two special points at zero $B$-field with $I_{DC}$ value set by the condition $v_{drift} =\pm s$ or (equivalently $I_{DC}=\pm 15.9$ $\mu$A for our situation). Following the insightful work of Dmitriev \textit{et al.} \cite{dmitriev2010}, magneto-resistance data presented in the form of two-dimensional plots in $B$ versus $I_{DC}$ space can be transformed to plots in $\epsilon_{ph}$  versus $\epsilon_{DC}$ space. When viewed in $\epsilon_{ph}$ versus $\epsilon_{DC}$ space, PIRO features track diagonal (anti-diagonal) lines with integer, or close to integer, value of $\epsilon_{DC}-\epsilon_{ph}$ ($\epsilon_{DC}+\epsilon_{ph}$) [see Fig. \ref{fig:phase} panels (a) and (d)]. The diagonal (anti-diagonal) running lines relate to the PIRO features in $B$ versus $I_{DC}$ space originating from the special point at zero $B$-field at positive (negative) DC current when $v_{drift} =+s$ ($v_{drift} =-s$). The ``sound barrier'' corresponds to the diagonal $\epsilon_{DC}-\epsilon_{ph}=0$ black dashed line- its twin anti-diagonal line for $\epsilon_{DC}+\epsilon_{ph}=0$ is not marked. Considering, for example, the $+\epsilon_{ph}$ and $+\epsilon_{DC}$ quadrant, $\epsilon_{DC}-\epsilon_{ph}<0$ ($\epsilon_{DC}-\epsilon_{ph}>0$) defines the subsonic (supersonic) regime. 

In their work \cite{dmitriev2010}, Dmitriev \textit{et al.} predict a subtle effect yet to be confirmed in experiment, namely a progressive $\pi/2$ change in the PIRO phase near the ``sound barrier'' on transitioning from the subsonic regime to the supersonic regime. Viewed along a vertical section through $\epsilon_{ph}$ versus $\epsilon_{DC}$ space, the expected signature of this effect is a steady shift in the PIRO maxima position when the magneto-resistance is plotted as a function of $\epsilon_{DC}-\epsilon_{ph}$ (in principle at a fixed value of $\epsilon_{ph}$): see bottom row panels in Fig. \ref{fig:phase}. 

Although we do not claim definitively to have observed the effect precisely as predicted, nonetheless we observe clear evidence for a phase change in the experimental data on transitioning across the ``sound barrier''.

Figure \ref{fig:phase}(a) shows the 3.9 K $\rho_{xx}^{*}$ data where PIRO features dominate in both the subsonic and supersonic regimes. The data is plotted in $\epsilon_{ph}$ versus $\epsilon_{DC}$ space. We observe the following: i. PIRO features generally track diagonal and anti-diagonal lines although a PIRO feature tracking such a line clearly ``wiggles'' along the line; ii. PIRO features at the ``sound barrier'' and (immediately) inside the supersonic regime are stronger than those in the subsonic regime and deep in the supersonic regime; and iii. the amplitude of PIRO features is typically strongest at or near points where $\epsilon_{ph}$ and $\epsilon_{DC}$ are both integers (and on very close inspection sometimes stronger near points where $\epsilon_{ph}$ and $\epsilon_{DC}$ are both half integers). Figure \ref{fig:phase}(d) shows $\rho_{xx}^{osc}$ calculated for $T=$ 3.9 K following the model in Ref. \cite{dmitriev2010} also plotted in $\epsilon_{ph}$ versus $\epsilon_{DC}$ space with the same input parameters as used to generate the plots in Fig. 2(d) and Fig. 3(bii). At a coarse level, the calculated plot in Fig. \ref{fig:phase}(d) and the plot of experimental data in Fig. \ref{fig:phase}(a) share many common features namely PIRO features running in a diagonal and anti-diagonal fashion, and enhanced amplitude at the ``sound barrier'' and in the supersonic regime compared to the subsonic regime. Nonetheless, the PIRO features in the calculated plot appear more straight in contrast to the ``wiggling" nature of the PIRO feature lines in the experimental data. Furthermore, the strong amplitude modulations of the PIRO features leading to the ``spoty'' pattern of maxima at or near points where $\epsilon_{ph}$ and $\epsilon_{DC}$ are either both integer or both half integers are very apparent in the calculation, whereas in the experimental data the amplitude modulations along the ``wiggling'' PIRO feature lines are more extended (or more ``streaky'') in appearance.

On close inspection, the predicted phase change effect is clear in the calculated plot in Fig. \ref{fig:phase}(d). Examining, for example, the $+\epsilon_{ph}$ and $+\epsilon_{DC}$ quadrant, there is clearly a PIRO line feature running along \emph{and exactly on} the diagonal $\epsilon_{DC}-\epsilon_{ph}=2$ line in the supersonic regime. On the other hand, in the subsonic regime there is clearly a PIRO line feature running parallel to \emph{but now below} the $\epsilon_{DC}-\epsilon_{ph}=-2$ line. The gradual shifting of the position of the PIRO line features over the range from $\epsilon_{DC}-\epsilon_{ph}=-2$ to $\epsilon_{DC}-\epsilon_{ph}=+2$ (essentially evident at any fixed value of $\epsilon_{ph}$) is due to the predicted $\pi/2$ phase change effect \cite{dmitriev2010}- see the five consecutive diagonal black dashed (guide) lines and the five related PIRO line features in panel (d). We stress the phase change is continuous and not abrupt across the ``sound barrier'', although for our situation, and the choice of parameters in the model calculations in Ref. \cite{dmitriev2010}, the phase change in practical terms is only readily apparent in the $\epsilon_{DC}-\epsilon_{ph} \in  [-1, +1]$ window and the phase has saturated for $\epsilon_{DC}-\epsilon_{ph}=\pm2$. Note also that the sum of the absolute values of the phase shift for the PIRO line features tracking any particular diagonal black line in the $+\epsilon_{ph}$ and $+\epsilon_{DC}$ quadrant \emph{and} in the $-\epsilon_{ph}$ and $-\epsilon_{DC}$ quadrant is $\pi/2$. For example, relative to the $\epsilon_{DC}-\epsilon_{ph}=+2$ diagonal line, the \emph{supersonic} PIRO line feature in the $+\epsilon_{ph}$ and $+\epsilon_{DC}$ quadrant is not shifted whereas the (now) \emph{subsonic} PIRO line feature in the $-\epsilon_{ph}$ and $-\epsilon_{DC}$ quadrant is (maximally) shifted by $\pi/2$. At the ``sound barrier'' (along the $\epsilon_{DC}-\epsilon_{ph}=0$ diagonal line), the PIRO line features in \emph{both} the $+\epsilon_{ph}$ and $+\epsilon_{DC}$ and $-\epsilon_{ph}$ and $-\epsilon_{DC}$ quadrants are shifted equally by $\pi/4$.    

Do we see evidence for the predicted phase change effect in the experimental data? Because the experimentally observed PIROs appear in fine detail as ``wiggly" line features with delocalized ``streaky'' amplitude modulations rather than straight line features with localized ``spotty'' amplitude modulations, as in the calculation, it could be difficult to discern the phase change effect by looking at PIRO features just along a vertical section through Fig. \ref{fig:phase}(a) at a single value of $\epsilon_{ph}$. Therefore, we examine the average value of $\rho_{xx}^{*}$ over a range of $\epsilon_{ph}$ values. Figures \ref{fig:phase}(b) and (c) show respectively traces of $\rho_{xx}^{*}$ averaged over $\epsilon_{ph} \in [-4, -2]$ and $\epsilon_{ph} \in [+2, +4]$ versus $\epsilon_{DC}-\epsilon_{ph}$. In the $+\epsilon_{ph}$ and $+\epsilon_{DC}$ quadrant [panel (c)], the PIRO maximum at $\epsilon_{DC}-\epsilon_{ph}=+2$ in the supersonic regime is not shifted away from its expected value, whereas the PIRO maximum expected at $\epsilon_{DC}-\epsilon_{ph}=-1$ in the subsonic regime appears shifted towards $\epsilon_{DC}-\epsilon_{ph} \sim-1.25$. The situation is inverted in the $-\epsilon_{ph}$ and $-\epsilon_{DC}$ quadrant [panel (b)]. The (weak) PIRO maximum at $\epsilon_{DC}-\epsilon_{ph} \sim-2$ in the supersonic regime is close to its expected value, whereas the PIRO maximum expected at $\epsilon_{DC}-\epsilon_{ph}=+1$ in the subsonic regime appears shifted towards $\epsilon_{DC}-\epsilon_{ph} \sim+1.25$. The PIRO maximum closest to the ``sound barrier'' ($\epsilon_{DC}-\epsilon_{ph}=0$) appears shifted in both quadrants. These empirical observations are consistent with those in the corresponding calculated traces in Figs. \ref{fig:phase}(e) and (f) following the theoretical description by Dmitriev \textit{et al.} where for consistency we have also averaged $\rho_{xx}^{osc}$ over $\epsilon_{ph} \in [-4, -2]$ and $\epsilon_{ph} \in [+2, +4]$ respectively. Nonetheless, in the experimental data, the PIRO maxima wash out deep in the supersonic regime (at high current), and the cut-off in the subsonic regime is set by the choice of the $\epsilon_{ph}$ averaging window width and its position. PIRO features also wash out at high $\epsilon_{ph}$ (small B) limited by $\tau_{q}$, and at small $\epsilon_{ph}$ (large B) the quantum Hall effect sets in a some point.

We should note that in Ref. \cite{hatke2011piro} Hatke \textit{et al.} discuss the phase of PIROs at \emph{zero DC current}, i.e., at $\epsilon_{DC}=0$. They report a non-zero phase $\sim$$\pi /4$ in the PIRO peaks. A constant non-zero phase of $\pi /4$ at \emph{zero DC current} is expected according to the theory in Ref. \cite{dmitriev2010} except when $\epsilon_{ph}$ approaches zero. We strongly stress that the near constant $\pi/4$ phase of PIROs for $\epsilon_{DC}=0$ under close to equilibrium conditions reported in Ref. \cite{hatke2011piro} and the $\pi/2$ phase shift across the ``sound barrier'' for diagonal, and anti-diagonal features under non-equilibrium conditions in Fig. \ref{fig:phase} that we discuss, are both consistent with the predictions of the same theory in Ref. \cite{dmitriev2010}.

%% file: main.bbl
\begin{thebibliography}{79}%
\makeatletter
\providecommand \@ifxundefined [1]{%
 \@ifx{#1\undefined}
}%
\providecommand \@ifnum [1]{%
 \ifnum #1\expandafter \@firstoftwo
 \else \expandafter \@secondoftwo
 \fi
}%
\providecommand \@ifx [1]{%
 \ifx #1\expandafter \@firstoftwo
 \else \expandafter \@secondoftwo
 \fi
}%
\providecommand \natexlab [1]{#1}%
\providecommand \enquote  [1]{``#1''}%
\providecommand \bibnamefont  [1]{#1}%
\providecommand \bibfnamefont [1]{#1}%
\providecommand \citenamefont [1]{#1}%
\providecommand \href@noop [0]{\@secondoftwo}%
\providecommand \href [0]{\begingroup \@sanitize@url \@href}%
\providecommand \@href[1]{\@@startlink{#1}\@@href}%
\providecommand \@@href[1]{\endgroup#1\@@endlink}%
\providecommand \@sanitize@url [0]{\catcode `\\12\catcode `\$12\catcode `\&12\catcode `\#12\catcode `\^12\catcode `\_12\catcode `\%12\relax}%
\providecommand \@@startlink[1]{}%
\providecommand \@@endlink[0]{}%
\providecommand \url  [0]{\begingroup\@sanitize@url \@url }%
\providecommand \@url [1]{\endgroup\@href {#1}{\urlprefix }}%
\providecommand \urlprefix  [0]{URL }%
\providecommand \Eprint [0]{\href }%
\providecommand \doibase [0]{https://doi.org/}%
\providecommand \selectlanguage [0]{\@gobble}%
\providecommand \bibinfo  [0]{\@secondoftwo}%
\providecommand \bibfield  [0]{\@secondoftwo}%
\providecommand \translation [1]{[#1]}%
\providecommand \BibitemOpen [0]{}%
\providecommand \bibitemStop [0]{}%
\providecommand \bibitemNoStop [0]{.\EOS\space}%
\providecommand \EOS [0]{\spacefactor3000\relax}%
\providecommand \BibitemShut  [1]{\csname bibitem#1\endcsname}%
\let\auto@bib@innerbib\@empty
\bibitem [{\citenamefont {Kaganov}\ \emph {et~al.}(1957)\citenamefont {Kaganov}, \citenamefont {Lifshitz},\ and\ \citenamefont {Tanatarov}}]{kaganov1957}%
  \BibitemOpen
  \bibfield  {author} {\bibinfo {author} {\bibfnamefont {M.~I.}\ \bibnamefont {Kaganov}}, \bibinfo {author} {\bibfnamefont {I.~M.}\ \bibnamefont {Lifshitz}},\ and\ \bibinfo {author} {\bibfnamefont {L.~V.}\ \bibnamefont {Tanatarov}},\ }\href@noop {} {\bibfield  {journal} {\bibinfo  {journal} {Sov. Phys. JETP}\ }\textbf {\bibinfo {volume} {4}},\ \bibinfo {pages} {173} (\bibinfo {year} {1957})}\BibitemShut {NoStop}%
\bibitem [{\citenamefont {Hutson}\ \emph {et~al.}(1961)\citenamefont {Hutson}, \citenamefont {McFee},\ and\ \citenamefont {White}}]{Hutson1961}%
  \BibitemOpen
  \bibfield  {author} {\bibinfo {author} {\bibfnamefont {A.~R.}\ \bibnamefont {Hutson}}, \bibinfo {author} {\bibfnamefont {J.~H.}\ \bibnamefont {McFee}},\ and\ \bibinfo {author} {\bibfnamefont {D.~L.}\ \bibnamefont {White}},\ }\href {https://doi.org/10.1103/PhysRevLett.7.237} {\bibfield  {journal} {\bibinfo  {journal} {Phys. Rev. Lett.}\ }\textbf {\bibinfo {volume} {7}},\ \bibinfo {pages} {237} (\bibinfo {year} {1961})}\BibitemShut {NoStop}%
\bibitem [{\citenamefont {Esaki}(1962)}]{Esaki1962}%
  \BibitemOpen
  \bibfield  {author} {\bibinfo {author} {\bibfnamefont {L.}~\bibnamefont {Esaki}},\ }\href {https://doi.org/10.1103/PhysRevLett.8.4} {\bibfield  {journal} {\bibinfo  {journal} {Phys. Rev. Lett.}\ }\textbf {\bibinfo {volume} {8}},\ \bibinfo {pages} {4} (\bibinfo {year} {1962})}\BibitemShut {NoStop}%
\bibitem [{\citenamefont {Smith}(1962)}]{Smith1962}%
  \BibitemOpen
  \bibfield  {author} {\bibinfo {author} {\bibfnamefont {R.~W.}\ \bibnamefont {Smith}},\ }\href {https://doi.org/10.1103/PhysRevLett.9.87} {\bibfield  {journal} {\bibinfo  {journal} {Phys. Rev. Lett.}\ }\textbf {\bibinfo {volume} {9}},\ \bibinfo {pages} {87} (\bibinfo {year} {1962})}\BibitemShut {NoStop}%
\bibitem [{\citenamefont {Wang}(1962)}]{Wang1962}%
  \BibitemOpen
  \bibfield  {author} {\bibinfo {author} {\bibfnamefont {W.-C.}\ \bibnamefont {Wang}},\ }\href {https://doi.org/10.1103/PhysRevLett.9.443} {\bibfield  {journal} {\bibinfo  {journal} {Phys. Rev. Lett.}\ }\textbf {\bibinfo {volume} {9}},\ \bibinfo {pages} {443} (\bibinfo {year} {1962})}\BibitemShut {NoStop}%
\bibitem [{\citenamefont {Spector}(1963)}]{Spector1963}%
  \BibitemOpen
  \bibfield  {author} {\bibinfo {author} {\bibfnamefont {H.~N.}\ \bibnamefont {Spector}},\ }\href {https://doi.org/10.1103/PhysRev.131.2512} {\bibfield  {journal} {\bibinfo  {journal} {Phys. Rev.}\ }\textbf {\bibinfo {volume} {131}},\ \bibinfo {pages} {2512} (\bibinfo {year} {1963})}\BibitemShut {NoStop}%
\bibitem [{\citenamefont {Eckstein}(1963)}]{Eckstein1963}%
  \BibitemOpen
  \bibfield  {author} {\bibinfo {author} {\bibfnamefont {S.~G.}\ \bibnamefont {Eckstein}},\ }\href {https://doi.org/10.1103/PhysRev.131.1087} {\bibfield  {journal} {\bibinfo  {journal} {Phys. Rev.}\ }\textbf {\bibinfo {volume} {131}},\ \bibinfo {pages} {1087} (\bibinfo {year} {1963})}\BibitemShut {NoStop}%
\bibitem [{\citenamefont {Prohofsky}\ and\ \citenamefont {Krumhansl}(1964)}]{Prohofsky1964}%
  \BibitemOpen
  \bibfield  {author} {\bibinfo {author} {\bibfnamefont {E.~W.}\ \bibnamefont {Prohofsky}}\ and\ \bibinfo {author} {\bibfnamefont {J.~A.}\ \bibnamefont {Krumhansl}},\ }\href {https://doi.org/10.1103/PhysRev.133.A1403} {\bibfield  {journal} {\bibinfo  {journal} {Phys. Rev.}\ }\textbf {\bibinfo {volume} {133}},\ \bibinfo {pages} {A1403} (\bibinfo {year} {1964})}\BibitemShut {NoStop}%
\bibitem [{\citenamefont {Zylbersztejn}(1967)}]{Zylber1967}%
  \BibitemOpen
  \bibfield  {author} {\bibinfo {author} {\bibfnamefont {A.}~\bibnamefont {Zylbersztejn}},\ }\href {https://doi.org/10.1103/PhysRevLett.19.838} {\bibfield  {journal} {\bibinfo  {journal} {Phys. Rev. Lett.}\ }\textbf {\bibinfo {volume} {19}},\ \bibinfo {pages} {838} (\bibinfo {year} {1967})}\BibitemShut {NoStop}%
\bibitem [{\citenamefont {Beardsley}\ \emph {et~al.}(2010)\citenamefont {Beardsley}, \citenamefont {Akimov}, \citenamefont {Henini},\ and\ \citenamefont {Kent}}]{Beardsley2010}%
  \BibitemOpen
  \bibfield  {author} {\bibinfo {author} {\bibfnamefont {R.~P.}\ \bibnamefont {Beardsley}}, \bibinfo {author} {\bibfnamefont {A.~V.}\ \bibnamefont {Akimov}}, \bibinfo {author} {\bibfnamefont {M.}~\bibnamefont {Henini}},\ and\ \bibinfo {author} {\bibfnamefont {A.~J.}\ \bibnamefont {Kent}},\ }\href {https://doi.org/10.1103/PhysRevLett.104.085501} {\bibfield  {journal} {\bibinfo  {journal} {Phys. Rev. Lett.}\ }\textbf {\bibinfo {volume} {104}},\ \bibinfo {pages} {085501} (\bibinfo {year} {2010})}\BibitemShut {NoStop}%
\bibitem [{\citenamefont {Maryam}\ \emph {et~al.}(2013)\citenamefont {Maryam}, \citenamefont {Akimov}, \citenamefont {Campion},\ and\ \citenamefont {Kent}}]{Maryam2013}%
  \BibitemOpen
  \bibfield  {author} {\bibinfo {author} {\bibfnamefont {W.}~\bibnamefont {Maryam}}, \bibinfo {author} {\bibfnamefont {A.~V.}\ \bibnamefont {Akimov}}, \bibinfo {author} {\bibfnamefont {R.}~\bibnamefont {Campion}},\ and\ \bibinfo {author} {\bibfnamefont {A.}~\bibnamefont {Kent}},\ }\href {https://doi.org/10.1038/ncomms3184} {\bibfield  {journal} {\bibinfo  {journal} {Nature Communications}\ }\textbf {\bibinfo {volume} {4}},\ \bibinfo {pages} {2184} (\bibinfo {year} {2013})}\BibitemShut {NoStop}%
\bibitem [{\citenamefont {Shinokita}\ \emph {et~al.}(2016)\citenamefont {Shinokita}, \citenamefont {Reimann}, \citenamefont {Woerner}, \citenamefont {Elsaesser}, \citenamefont {Hey},\ and\ \citenamefont {Flytzanis}}]{Shinokita2016}%
  \BibitemOpen
  \bibfield  {author} {\bibinfo {author} {\bibfnamefont {K.}~\bibnamefont {Shinokita}}, \bibinfo {author} {\bibfnamefont {K.}~\bibnamefont {Reimann}}, \bibinfo {author} {\bibfnamefont {M.}~\bibnamefont {Woerner}}, \bibinfo {author} {\bibfnamefont {T.}~\bibnamefont {Elsaesser}}, \bibinfo {author} {\bibfnamefont {R.}~\bibnamefont {Hey}},\ and\ \bibinfo {author} {\bibfnamefont {C.}~\bibnamefont {Flytzanis}},\ }\href {https://doi.org/10.1103/PhysRevLett.116.075504} {\bibfield  {journal} {\bibinfo  {journal} {Phys. Rev. Lett.}\ }\textbf {\bibinfo {volume} {116}},\ \bibinfo {pages} {075504} (\bibinfo {year} {2016})}\BibitemShut {NoStop}%
\bibitem [{\citenamefont {Barajas-Aguilar}\ \emph {et~al.}(2024)\citenamefont {Barajas-Aguilar}, \citenamefont {Zion}, \citenamefont {Sequeira}, \citenamefont {Barabas}, \citenamefont {Taniguchi}, \citenamefont {Watanabe}, \citenamefont {Barrett}, \citenamefont {Scaffidi},\ and\ \citenamefont {Sanchez-Yamagishi}}]{Barajas2024}%
  \BibitemOpen
  \bibfield  {author} {\bibinfo {author} {\bibfnamefont {A.~H.}\ \bibnamefont {Barajas-Aguilar}}, \bibinfo {author} {\bibfnamefont {J.}~\bibnamefont {Zion}}, \bibinfo {author} {\bibfnamefont {I.}~\bibnamefont {Sequeira}}, \bibinfo {author} {\bibfnamefont {A.~Z.}\ \bibnamefont {Barabas}}, \bibinfo {author} {\bibfnamefont {T.}~\bibnamefont {Taniguchi}}, \bibinfo {author} {\bibfnamefont {K.}~\bibnamefont {Watanabe}}, \bibinfo {author} {\bibfnamefont {E.~B.}\ \bibnamefont {Barrett}}, \bibinfo {author} {\bibfnamefont {T.}~\bibnamefont {Scaffidi}},\ and\ \bibinfo {author} {\bibfnamefont {J.~D.}\ \bibnamefont {Sanchez-Yamagishi}},\ }\href {https://doi.org/10.1038/s41467-024-46819-2} {\bibfield  {journal} {\bibinfo  {journal} {Nature Communications}\ }\textbf {\bibinfo {volume} {15}},\ \bibinfo {pages} {2550} (\bibinfo {year} {2024})}\BibitemShut {NoStop}%
\bibitem [{\citenamefont {Wendt}\ \emph {et~al.}(2025)\citenamefont {Wendt}, \citenamefont {Storey}, \citenamefont {Miller}, \citenamefont {Anderson}, \citenamefont {Chatterjee}, \citenamefont {Horrocks}, \citenamefont {Smith},\ and\ \citenamefont {Eichenfield}}]{Wendt2025}%
  \BibitemOpen
  \bibfield  {author} {\bibinfo {author} {\bibfnamefont {A.}~\bibnamefont {Wendt}}, \bibinfo {author} {\bibfnamefont {M.~J.}\ \bibnamefont {Storey}}, \bibinfo {author} {\bibfnamefont {M.}~\bibnamefont {Miller}}, \bibinfo {author} {\bibfnamefont {D.}~\bibnamefont {Anderson}}, \bibinfo {author} {\bibfnamefont {E.}~\bibnamefont {Chatterjee}}, \bibinfo {author} {\bibfnamefont {W.}~\bibnamefont {Horrocks}}, \bibinfo {author} {\bibfnamefont {B.}~\bibnamefont {Smith}},\ and\ \bibinfo {author} {\bibfnamefont {L.~H.~M.}\ \bibnamefont {Eichenfield}},\ }\href@noop {} {\bibfield  {journal} {\bibinfo  {journal} {arXiv preprint arXiv:2505.14385}\ } (\bibinfo {year} {2025})}\BibitemShut {NoStop}%
\bibitem [{\citenamefont {Garc\'{\i}a~de Abajo}(2010)}]{Abajo2010}%
  \BibitemOpen
  \bibfield  {author} {\bibinfo {author} {\bibfnamefont {F.~J.}\ \bibnamefont {Garc\'{\i}a~de Abajo}},\ }\href {https://doi.org/10.1103/RevModPhys.82.209} {\bibfield  {journal} {\bibinfo  {journal} {Rev. Mod. Phys.}\ }\textbf {\bibinfo {volume} {82}},\ \bibinfo {pages} {209} (\bibinfo {year} {2010})}\BibitemShut {NoStop}%
\bibitem [{\citenamefont {Andersen}\ \emph {et~al.}(2019)\citenamefont {Andersen}, \citenamefont {Dwyer}, \citenamefont {Sanchez-Yamagishi}, \citenamefont {Rodriguez-Nieva}, \citenamefont {Agarwal}, \citenamefont {Watanabe}, \citenamefont {Taniguchi}, \citenamefont {Demler}, \citenamefont {Kim}, \citenamefont {Park} \emph {et~al.}}]{andersen2019electron}%
  \BibitemOpen
  \bibfield  {author} {\bibinfo {author} {\bibfnamefont {T.~I.}\ \bibnamefont {Andersen}}, \bibinfo {author} {\bibfnamefont {B.~L.}\ \bibnamefont {Dwyer}}, \bibinfo {author} {\bibfnamefont {J.~D.}\ \bibnamefont {Sanchez-Yamagishi}}, \bibinfo {author} {\bibfnamefont {J.~F.}\ \bibnamefont {Rodriguez-Nieva}}, \bibinfo {author} {\bibfnamefont {K.}~\bibnamefont {Agarwal}}, \bibinfo {author} {\bibfnamefont {K.}~\bibnamefont {Watanabe}}, \bibinfo {author} {\bibfnamefont {T.}~\bibnamefont {Taniguchi}}, \bibinfo {author} {\bibfnamefont {E.~A.}\ \bibnamefont {Demler}}, \bibinfo {author} {\bibfnamefont {P.}~\bibnamefont {Kim}}, \bibinfo {author} {\bibfnamefont {H.}~\bibnamefont {Park}}, \emph {et~al.},\ }\href {https://doi.org/10.1126/science.aaw2104} {\bibfield  {journal} {\bibinfo  {journal} {Science}\ }\textbf {\bibinfo {volume} {364}},\ \bibinfo {pages} {154} (\bibinfo {year} {2019})}\BibitemShut {NoStop}%
\bibitem [{\citenamefont {Greenaway}\ \emph {et~al.}(2021)\citenamefont {Greenaway}, \citenamefont {Kumaravadivel}, \citenamefont {Wengraf}, \citenamefont {Ponomarenko}, \citenamefont {Berdyugin}, \citenamefont {Li}, \citenamefont {Edgar}, \citenamefont {Kumar}, \citenamefont {Geim},\ and\ \citenamefont {Eaves}}]{greenaway2021}%
  \BibitemOpen
  \bibfield  {author} {\bibinfo {author} {\bibfnamefont {M.}~\bibnamefont {Greenaway}}, \bibinfo {author} {\bibfnamefont {P.}~\bibnamefont {Kumaravadivel}}, \bibinfo {author} {\bibfnamefont {J.}~\bibnamefont {Wengraf}}, \bibinfo {author} {\bibfnamefont {L.}~\bibnamefont {Ponomarenko}}, \bibinfo {author} {\bibfnamefont {A.}~\bibnamefont {Berdyugin}}, \bibinfo {author} {\bibfnamefont {J.}~\bibnamefont {Li}}, \bibinfo {author} {\bibfnamefont {J.}~\bibnamefont {Edgar}}, \bibinfo {author} {\bibfnamefont {R.~K.}\ \bibnamefont {Kumar}}, \bibinfo {author} {\bibfnamefont {A.}~\bibnamefont {Geim}},\ and\ \bibinfo {author} {\bibfnamefont {L.}~\bibnamefont {Eaves}},\ }\href@noop {} {\bibfield  {journal} {\bibinfo  {journal} {Nature Communications}\ }\textbf {\bibinfo {volume} {12}},\ \bibinfo {pages} {6392} (\bibinfo {year} {2021})}\BibitemShut {NoStop}%
\bibitem [{\citenamefont {Hu}\ \emph {et~al.}(2022)\citenamefont {Hu}, \citenamefont {Lin}, \citenamefont {Liu}, \citenamefont {Chen}, \citenamefont {Zhang},\ and\ \citenamefont {Luo}}]{Hu2022}%
  \BibitemOpen
  \bibfield  {author} {\bibinfo {author} {\bibfnamefont {H.}~\bibnamefont {Hu}}, \bibinfo {author} {\bibfnamefont {X.}~\bibnamefont {Lin}}, \bibinfo {author} {\bibfnamefont {D.}~\bibnamefont {Liu}}, \bibinfo {author} {\bibfnamefont {H.}~\bibnamefont {Chen}}, \bibinfo {author} {\bibfnamefont {B.}~\bibnamefont {Zhang}},\ and\ \bibinfo {author} {\bibfnamefont {Y.}~\bibnamefont {Luo}},\ }\href {https://doi.org/10.1002/advs.202200538} {\bibfield  {journal} {\bibinfo  {journal} {Advanced Science}\ }\textbf {\bibinfo {volume} {9}},\ \bibinfo {pages} {2200538} (\bibinfo {year} {2022})}\BibitemShut {NoStop}%
\bibitem [{\citenamefont {Geurs}\ \emph {et~al.}(2025)\citenamefont {Geurs}, \citenamefont {Webb}, \citenamefont {Guo}, \citenamefont {Keren}, \citenamefont {Farrell}, \citenamefont {Xu}, \citenamefont {Watanabe}, \citenamefont {Taniguchi}, \citenamefont {Basov}, \citenamefont {Hone}, \citenamefont {Lucas}, \citenamefont {Pasupathy},\ and\ \citenamefont {Dean}}]{Geurs2025}%
  \BibitemOpen
  \bibfield  {author} {\bibinfo {author} {\bibfnamefont {J.}~\bibnamefont {Geurs}}, \bibinfo {author} {\bibfnamefont {T.~A.}\ \bibnamefont {Webb}}, \bibinfo {author} {\bibfnamefont {Y.}~\bibnamefont {Guo}}, \bibinfo {author} {\bibfnamefont {I.}~\bibnamefont {Keren}}, \bibinfo {author} {\bibfnamefont {J.~H.}\ \bibnamefont {Farrell}}, \bibinfo {author} {\bibfnamefont {J.}~\bibnamefont {Xu}}, \bibinfo {author} {\bibfnamefont {K.}~\bibnamefont {Watanabe}}, \bibinfo {author} {\bibfnamefont {T.}~\bibnamefont {Taniguchi}}, \bibinfo {author} {\bibfnamefont {D.~N.}\ \bibnamefont {Basov}}, \bibinfo {author} {\bibfnamefont {J.}~\bibnamefont {Hone}}, \bibinfo {author} {\bibfnamefont {A.}~\bibnamefont {Lucas}}, \bibinfo {author} {\bibfnamefont {A.}~\bibnamefont {Pasupathy}},\ and\ \bibinfo {author} {\bibfnamefont {C.~R.}\ \bibnamefont {Dean}},\ }\href {https://arxiv.org/abs/2509.16321} {\bibinfo {title} {Supersonic flow and hydraulic jump in an electronic de laval nozzle}} (\bibinfo {year} {2025}),\ \Eprint
  {https://arxiv.org/abs/2509.16321} {arXiv:2509.16321 [cond-mat.mes-hall]} \BibitemShut {NoStop}%
\bibitem [{\citenamefont {Dong}\ \emph {et~al.}(2025)\citenamefont {Dong}, \citenamefont {Sun}, \citenamefont {Phinney}, \citenamefont {Sun}, \citenamefont {Andersen}, \citenamefont {Xiong}, \citenamefont {Shao}, \citenamefont {Zhang}, \citenamefont {Rikhter}, \citenamefont {Liu} \emph {et~al.}}]{dong2025current}%
  \BibitemOpen
  \bibfield  {author} {\bibinfo {author} {\bibfnamefont {Y.}~\bibnamefont {Dong}}, \bibinfo {author} {\bibfnamefont {Z.}~\bibnamefont {Sun}}, \bibinfo {author} {\bibfnamefont {I.}~\bibnamefont {Phinney}}, \bibinfo {author} {\bibfnamefont {D.}~\bibnamefont {Sun}}, \bibinfo {author} {\bibfnamefont {T.}~\bibnamefont {Andersen}}, \bibinfo {author} {\bibfnamefont {L.}~\bibnamefont {Xiong}}, \bibinfo {author} {\bibfnamefont {Y.}~\bibnamefont {Shao}}, \bibinfo {author} {\bibfnamefont {S.}~\bibnamefont {Zhang}}, \bibinfo {author} {\bibfnamefont {A.}~\bibnamefont {Rikhter}}, \bibinfo {author} {\bibfnamefont {S.}~\bibnamefont {Liu}}, \emph {et~al.},\ }\href {https://doi.org/10.1038/s41467-025-58953-6} {\bibfield  {journal} {\bibinfo  {journal} {Nature Communications}\ }\textbf {\bibinfo {volume} {16}},\ \bibinfo {pages} {3861} (\bibinfo {year} {2025})}\BibitemShut {NoStop}%
\bibitem [{\citenamefont {Vitkalov}(2009)}]{vitkalov2009NL}%
  \BibitemOpen
  \bibfield  {author} {\bibinfo {author} {\bibfnamefont {S.}~\bibnamefont {Vitkalov}},\ }\href {https://doi.org/https://doi.org/10.1142/S0217979209054090} {\bibfield  {journal} {\bibinfo  {journal} {Int. J. Mod. Phys. B}\ }\textbf {\bibinfo {volume} {23}},\ \bibinfo {pages} {4727} (\bibinfo {year} {2009})}\BibitemShut {NoStop}%
\bibitem [{\citenamefont {Dmitriev}(2011)}]{dmitriev2011}%
  \BibitemOpen
  \bibfield  {author} {\bibinfo {author} {\bibfnamefont {I.~A.}\ \bibnamefont {Dmitriev}},\ }\href {https://doi.org/10.1088/1742-6596/334/1/012015} {\bibfield  {journal} {\bibinfo  {journal} {J. Phys. Conf. Ser.}\ }\textbf {\bibinfo {volume} {334}},\ \bibinfo {pages} {012015} (\bibinfo {year} {2011})}\BibitemShut {NoStop}%
\bibitem [{\citenamefont {Dmitriev}\ \emph {et~al.}(2012)\citenamefont {Dmitriev}, \citenamefont {Mirlin}, \citenamefont {Polyakov},\ and\ \citenamefont {Zudov}}]{dmitriev2012review}%
  \BibitemOpen
  \bibfield  {author} {\bibinfo {author} {\bibfnamefont {I.~A.}\ \bibnamefont {Dmitriev}}, \bibinfo {author} {\bibfnamefont {A.~D.}\ \bibnamefont {Mirlin}}, \bibinfo {author} {\bibfnamefont {D.~G.}\ \bibnamefont {Polyakov}},\ and\ \bibinfo {author} {\bibfnamefont {M.~A.}\ \bibnamefont {Zudov}},\ }\href {https://doi.org/10.1103/RevModPhys.84.1709} {\bibfield  {journal} {\bibinfo  {journal} {Rev. Mod. Phys.}\ }\textbf {\bibinfo {volume} {84}},\ \bibinfo {pages} {1709} (\bibinfo {year} {2012})}\BibitemShut {NoStop}%
\bibitem [{epo()}]{epoint}%
  \BibitemOpen
  \href@noop {} {}\bibinfo {note} {Under equilibrium conditions, PIROs are rapidly diminished at very low temperature since, for scattering processes involving phonon absorption, the ambient phonon population in the lattice bath is strongly suppressed, and, for processes involving phonon emission, there are essentially no empty electronic states in Landau levels below the Fermi energy for carriers to drop into \cite{dmitriev2012review}.}\BibitemShut {Stop}%
\bibitem [{\citenamefont {Zudov}\ \emph {et~al.}(2001)\citenamefont {Zudov}, \citenamefont {Ponomarev}, \citenamefont {Efros}, \citenamefont {Du}, \citenamefont {Simmons},\ and\ \citenamefont {Reno}}]{zudov2001}%
  \BibitemOpen
  \bibfield  {author} {\bibinfo {author} {\bibfnamefont {M.~A.}\ \bibnamefont {Zudov}}, \bibinfo {author} {\bibfnamefont {I.~V.}\ \bibnamefont {Ponomarev}}, \bibinfo {author} {\bibfnamefont {A.~L.}\ \bibnamefont {Efros}}, \bibinfo {author} {\bibfnamefont {R.~R.}\ \bibnamefont {Du}}, \bibinfo {author} {\bibfnamefont {J.~A.}\ \bibnamefont {Simmons}},\ and\ \bibinfo {author} {\bibfnamefont {J.~L.}\ \bibnamefont {Reno}},\ }\href {https://doi.org/10.1103/physrevlett.86.3614} {\bibfield  {journal} {\bibinfo  {journal} {Phys. Rev. Lett.}\ }\textbf {\bibinfo {volume} {86}},\ \bibinfo {pages} {3614} (\bibinfo {year} {2001})}\BibitemShut {NoStop}%
\bibitem [{\citenamefont {Yang}\ \emph {et~al.}(2002{\natexlab{a}})\citenamefont {Yang}, \citenamefont {Zudov}, \citenamefont {Zhang}, \citenamefont {Du}, \citenamefont {Simmons},\ and\ \citenamefont {Reno}}]{yang2002A}%
  \BibitemOpen
  \bibfield  {author} {\bibinfo {author} {\bibfnamefont {C.}~\bibnamefont {Yang}}, \bibinfo {author} {\bibfnamefont {M.}~\bibnamefont {Zudov}}, \bibinfo {author} {\bibfnamefont {J.}~\bibnamefont {Zhang}}, \bibinfo {author} {\bibfnamefont {R.}~\bibnamefont {Du}}, \bibinfo {author} {\bibfnamefont {J.}~\bibnamefont {Simmons}},\ and\ \bibinfo {author} {\bibfnamefont {J.}~\bibnamefont {Reno}},\ }\href@noop {} {\bibfield  {journal} {\bibinfo  {journal} {Physica E: Low-dimensional Systems and Nanostructures}\ }\textbf {\bibinfo {volume} {12}},\ \bibinfo {pages} {443} (\bibinfo {year} {2002}{\natexlab{a}})}\BibitemShut {NoStop}%
\bibitem [{\citenamefont {Bykov}\ \emph {et~al.}(2005{\natexlab{a}})\citenamefont {Bykov}, \citenamefont {Kalagin},\ and\ \citenamefont {Bakarov}}]{bykov2005MP}%
  \BibitemOpen
  \bibfield  {author} {\bibinfo {author} {\bibfnamefont {A.~A.}\ \bibnamefont {Bykov}}, \bibinfo {author} {\bibfnamefont {A.}~\bibnamefont {Kalagin}},\ and\ \bibinfo {author} {\bibfnamefont {A.~K.}\ \bibnamefont {Bakarov}},\ }\href@noop {} {\bibfield  {journal} {\bibinfo  {journal} {Journal of Experimental and Theoretical Physics Letters}\ }\textbf {\bibinfo {volume} {81}},\ \bibinfo {pages} {523} (\bibinfo {year} {2005}{\natexlab{a}})}\BibitemShut {NoStop}%
\bibitem [{\citenamefont {Zhang}\ \emph {et~al.}(2004)\citenamefont {Zhang}, \citenamefont {Lyo}, \citenamefont {Du}, \citenamefont {Simmons},\ and\ \citenamefont {Reno}}]{Zhang2004XYZ}%
  \BibitemOpen
  \bibfield  {author} {\bibinfo {author} {\bibfnamefont {J.}~\bibnamefont {Zhang}}, \bibinfo {author} {\bibfnamefont {S.~K.}\ \bibnamefont {Lyo}}, \bibinfo {author} {\bibfnamefont {R.~R.}\ \bibnamefont {Du}}, \bibinfo {author} {\bibfnamefont {J.~A.}\ \bibnamefont {Simmons}},\ and\ \bibinfo {author} {\bibfnamefont {J.~L.}\ \bibnamefont {Reno}},\ }\href {https://doi.org/10.1103/PhysRevLett.92.156802} {\bibfield  {journal} {\bibinfo  {journal} {Phys. Rev. Lett.}\ }\textbf {\bibinfo {volume} {92}},\ \bibinfo {pages} {156802} (\bibinfo {year} {2004})}\BibitemShut {NoStop}%
\bibitem [{\citenamefont {Hatke}\ \emph {et~al.}(2009{\natexlab{a}})\citenamefont {Hatke}, \citenamefont {Zudov}, \citenamefont {Pfeiffer},\ and\ \citenamefont {West}}]{hatke2009phonon}%
  \BibitemOpen
  \bibfield  {author} {\bibinfo {author} {\bibfnamefont {A.~T.}\ \bibnamefont {Hatke}}, \bibinfo {author} {\bibfnamefont {M.~A.}\ \bibnamefont {Zudov}}, \bibinfo {author} {\bibfnamefont {L.~N.}\ \bibnamefont {Pfeiffer}},\ and\ \bibinfo {author} {\bibfnamefont {K.~W.}\ \bibnamefont {West}},\ }\href {https://doi.org/10.1103/PhysRevLett.102.086808} {\bibfield  {journal} {\bibinfo  {journal} {Phys. Rev. Lett.}\ }\textbf {\bibinfo {volume} {102}},\ \bibinfo {pages} {086808} (\bibinfo {year} {2009}{\natexlab{a}})}\BibitemShut {NoStop}%
\bibitem [{\citenamefont {Bykov}\ and\ \citenamefont {Goran}(2009)}]{Bykov2009}%
  \BibitemOpen
  \bibfield  {author} {\bibinfo {author} {\bibfnamefont {A.~A.}\ \bibnamefont {Bykov}}\ and\ \bibinfo {author} {\bibfnamefont {A.~V.}\ \bibnamefont {Goran}},\ }\href@noop {} {\bibfield  {journal} {\bibinfo  {journal} {JETP Letters}\ }\textbf {\bibinfo {volume} {90}},\ \bibinfo {pages} {578} (\bibinfo {year} {2009})}\BibitemShut {NoStop}%
\bibitem [{\citenamefont {Bykov}\ \emph {et~al.}(2010)\citenamefont {Bykov}, \citenamefont {Goran},\ and\ \citenamefont {Vitkalov}}]{bykov2010}%
  \BibitemOpen
  \bibfield  {author} {\bibinfo {author} {\bibfnamefont {A.~A.}\ \bibnamefont {Bykov}}, \bibinfo {author} {\bibfnamefont {A.~V.}\ \bibnamefont {Goran}},\ and\ \bibinfo {author} {\bibfnamefont {S.~A.}\ \bibnamefont {Vitkalov}},\ }\href {https://doi.org/10.1103/PhysRevB.81.155322} {\bibfield  {journal} {\bibinfo  {journal} {Phys. Rev. B}\ }\textbf {\bibinfo {volume} {81}},\ \bibinfo {pages} {155322} (\bibinfo {year} {2010})}\BibitemShut {NoStop}%
\bibitem [{\citenamefont {Hatke}\ \emph {et~al.}(2010)\citenamefont {Hatke}, \citenamefont {Chiang}, \citenamefont {Zudov}, \citenamefont {Pfeiffer},\ and\ \citenamefont {West}}]{Hatke2010}%
  \BibitemOpen
  \bibfield  {author} {\bibinfo {author} {\bibfnamefont {A.~T.}\ \bibnamefont {Hatke}}, \bibinfo {author} {\bibfnamefont {H.-S.}\ \bibnamefont {Chiang}}, \bibinfo {author} {\bibfnamefont {M.~A.}\ \bibnamefont {Zudov}}, \bibinfo {author} {\bibfnamefont {L.~N.}\ \bibnamefont {Pfeiffer}},\ and\ \bibinfo {author} {\bibfnamefont {K.~W.}\ \bibnamefont {West}},\ }\href {https://doi.org/10.1103/PhysRevB.82.041304} {\bibfield  {journal} {\bibinfo  {journal} {Phys. Rev. B}\ }\textbf {\bibinfo {volume} {82}},\ \bibinfo {pages} {041304} (\bibinfo {year} {2010})}\BibitemShut {NoStop}%
\bibitem [{\citenamefont {Hatke}\ \emph {et~al.}(2011{\natexlab{a}})\citenamefont {Hatke}, \citenamefont {Zudov}, \citenamefont {Pfeiffer},\ and\ \citenamefont {West}}]{hatke2011piro}%
  \BibitemOpen
  \bibfield  {author} {\bibinfo {author} {\bibfnamefont {A.~T.}\ \bibnamefont {Hatke}}, \bibinfo {author} {\bibfnamefont {M.~A.}\ \bibnamefont {Zudov}}, \bibinfo {author} {\bibfnamefont {L.~N.}\ \bibnamefont {Pfeiffer}},\ and\ \bibinfo {author} {\bibfnamefont {K.~W.}\ \bibnamefont {West}},\ }\href {https://doi.org/10.1103/PhysRevB.84.121301} {\bibfield  {journal} {\bibinfo  {journal} {Phys. Rev. B}\ }\textbf {\bibinfo {volume} {84}},\ \bibinfo {pages} {121301(R)} (\bibinfo {year} {2011}{\natexlab{a}})}\BibitemShut {NoStop}%
\bibitem [{\citenamefont {Hatke}\ \emph {et~al.}(2012{\natexlab{a}})\citenamefont {Hatke}, \citenamefont {Zudov}, \citenamefont {Reno}, \citenamefont {Pfeiffer},\ and\ \citenamefont {West}}]{hatke2012giant}%
  \BibitemOpen
  \bibfield  {author} {\bibinfo {author} {\bibfnamefont {A.}~\bibnamefont {Hatke}}, \bibinfo {author} {\bibfnamefont {M.}~\bibnamefont {Zudov}}, \bibinfo {author} {\bibfnamefont {J.}~\bibnamefont {Reno}}, \bibinfo {author} {\bibfnamefont {L.}~\bibnamefont {Pfeiffer}},\ and\ \bibinfo {author} {\bibfnamefont {K.}~\bibnamefont {West}},\ }\href {https://doi.org/10.1103/PhysRevB.85.081304} {\bibfield  {journal} {\bibinfo  {journal} {Phys. Rev. B}\ }\textbf {\bibinfo {volume} {85}},\ \bibinfo {pages} {081304(R)} (\bibinfo {year} {2012}{\natexlab{a}})}\BibitemShut {NoStop}%
\bibitem [{\citenamefont {Kumaravadivel}\ \emph {et~al.}(2019)\citenamefont {Kumaravadivel}, \citenamefont {Greenaway}, \citenamefont {Perello}, \citenamefont {Berdyugin}, \citenamefont {Birkbeck}, \citenamefont {Wengraf}, \citenamefont {Liu}, \citenamefont {Edgar}, \citenamefont {Geim}, \citenamefont {Eaves},\ and\ \citenamefont {Krishna~Kumar}}]{kumara2019}%
  \BibitemOpen
  \bibfield  {author} {\bibinfo {author} {\bibfnamefont {P.}~\bibnamefont {Kumaravadivel}}, \bibinfo {author} {\bibfnamefont {M.}~\bibnamefont {Greenaway}}, \bibinfo {author} {\bibfnamefont {D.}~\bibnamefont {Perello}}, \bibinfo {author} {\bibfnamefont {A.}~\bibnamefont {Berdyugin}}, \bibinfo {author} {\bibfnamefont {J.}~\bibnamefont {Birkbeck}}, \bibinfo {author} {\bibfnamefont {J.}~\bibnamefont {Wengraf}}, \bibinfo {author} {\bibfnamefont {S.}~\bibnamefont {Liu}}, \bibinfo {author} {\bibfnamefont {J.}~\bibnamefont {Edgar}}, \bibinfo {author} {\bibfnamefont {A.}~\bibnamefont {Geim}}, \bibinfo {author} {\bibfnamefont {L.}~\bibnamefont {Eaves}},\ and\ \bibinfo {author} {\bibfnamefont {R.}~\bibnamefont {Krishna~Kumar}},\ }\href@noop {} {\bibfield  {journal} {\bibinfo  {journal} {Nature Communications}\ }\textbf {\bibinfo {volume} {10}},\ \bibinfo {pages} {3334} (\bibinfo {year} {2019})}\BibitemShut {NoStop}%
\bibitem [{\citenamefont {Greenaway}\ \emph {et~al.}(2019)\citenamefont {Greenaway}, \citenamefont {Krishna~Kumar}, \citenamefont {Kumaravadivel}, \citenamefont {Geim},\ and\ \citenamefont {Eaves}}]{Greenaway2019}%
  \BibitemOpen
  \bibfield  {author} {\bibinfo {author} {\bibfnamefont {M.~T.}\ \bibnamefont {Greenaway}}, \bibinfo {author} {\bibfnamefont {R.}~\bibnamefont {Krishna~Kumar}}, \bibinfo {author} {\bibfnamefont {P.}~\bibnamefont {Kumaravadivel}}, \bibinfo {author} {\bibfnamefont {A.~K.}\ \bibnamefont {Geim}},\ and\ \bibinfo {author} {\bibfnamefont {L.}~\bibnamefont {Eaves}},\ }\href {https://doi.org/10.1103/PhysRevB.100.155120} {\bibfield  {journal} {\bibinfo  {journal} {Phys. Rev. B}\ }\textbf {\bibinfo {volume} {100}},\ \bibinfo {pages} {155120} (\bibinfo {year} {2019})}\BibitemShut {NoStop}%
\bibitem [{\citenamefont {Zhang}\ \emph {et~al.}(2008)\citenamefont {Zhang}, \citenamefont {Zudov}, \citenamefont {Pfeiffer},\ and\ \citenamefont {West}}]{zhang2008}%
  \BibitemOpen
  \bibfield  {author} {\bibinfo {author} {\bibfnamefont {W.}~\bibnamefont {Zhang}}, \bibinfo {author} {\bibfnamefont {M.~A.}\ \bibnamefont {Zudov}}, \bibinfo {author} {\bibfnamefont {L.~N.}\ \bibnamefont {Pfeiffer}},\ and\ \bibinfo {author} {\bibfnamefont {K.~W.}\ \bibnamefont {West}},\ }\href {https://doi.org/10.1103/PhysRevLett.100.036805} {\bibfield  {journal} {\bibinfo  {journal} {Phys. Rev. Lett.}\ }\textbf {\bibinfo {volume} {100}},\ \bibinfo {pages} {036805} (\bibinfo {year} {2008})}\BibitemShut {NoStop}%
\bibitem [{sup()}]{supplemental}%
  \BibitemOpen
  \href@noop {} {}\bibinfo {note} {See Supplemental Material at WEBLINK for a description of the background subtraction from $\rho_{xx}$ to calculate $\rho_{xx}^{*}$, an analysis of the HIROs observed at 10 mK, a discussion on the resonant scattering processes involving acoustic phonons, a description of the fit of the subsonic 3.9 K data along with commentary on the estimated electron-phonon coupling constant, and commentary of the phase change effect transitioning from the subsonic regime to the supersonic regime.}\BibitemShut {Stop}%
\bibitem [{\citenamefont {Dmitriev}\ \emph {et~al.}(2010)\citenamefont {Dmitriev}, \citenamefont {Gellmann},\ and\ \citenamefont {Vavilov}}]{dmitriev2010}%
  \BibitemOpen
  \bibfield  {author} {\bibinfo {author} {\bibfnamefont {I.~A.}\ \bibnamefont {Dmitriev}}, \bibinfo {author} {\bibfnamefont {R.}~\bibnamefont {Gellmann}},\ and\ \bibinfo {author} {\bibfnamefont {M.~G.}\ \bibnamefont {Vavilov}},\ }\href {https://doi.org/10.1103/PhysRevB.82.201311} {\bibfield  {journal} {\bibinfo  {journal} {Phys. Rev. B}\ }\textbf {\bibinfo {volume} {82}},\ \bibinfo {pages} {201311(R)} (\bibinfo {year} {2010})}\BibitemShut {NoStop}%
\bibitem [{\citenamefont {Hatke}\ \emph {et~al.}(2015)\citenamefont {Hatke}, \citenamefont {Zudov},\ and\ \citenamefont {Reno}}]{hatke2015SAND}%
  \BibitemOpen
  \bibfield  {author} {\bibinfo {author} {\bibfnamefont {A.}~\bibnamefont {Hatke}}, \bibinfo {author} {\bibfnamefont {M.}~\bibnamefont {Zudov}},\ and\ \bibinfo {author} {\bibfnamefont {J.~L.}\ \bibnamefont {Reno}},\ }\href {https://www.osti.gov/servlets/purl/1427272} {\emph {\bibinfo {title} {Supersonic transport in GaAs/AlGaAs heterostructures}}},\ \bibinfo {type} {Tech. Rep.}\ (\bibinfo  {institution} {Sandia National Lab.(SNL-NM), Albuquerque, NM (United States)},\ \bibinfo {year} {2015})\BibitemShut {NoStop}%
\bibitem [{\citenamefont {Raichev}\ \emph {et~al.}(2017)\citenamefont {Raichev}, \citenamefont {Hatke}, \citenamefont {Zudov},\ and\ \citenamefont {Reno}}]{raichev2017}%
  \BibitemOpen
  \bibfield  {author} {\bibinfo {author} {\bibfnamefont {O.~E.}\ \bibnamefont {Raichev}}, \bibinfo {author} {\bibfnamefont {A.~T.}\ \bibnamefont {Hatke}}, \bibinfo {author} {\bibfnamefont {M.~A.}\ \bibnamefont {Zudov}},\ and\ \bibinfo {author} {\bibfnamefont {J.~L.}\ \bibnamefont {Reno}},\ }\href {https://doi.org/10.1103/PhysRevB.96.081407} {\bibfield  {journal} {\bibinfo  {journal} {Phys. Rev. B}\ }\textbf {\bibinfo {volume} {96}},\ \bibinfo {pages} {081407} (\bibinfo {year} {2017})}\BibitemShut {NoStop}%
\bibitem [{dpo()}]{dpoint}%
  \BibitemOpen
  \href@noop {} {}\bibinfo {note} {We note that although a single broad peak in the differential resistivity ascribed to the onset of supersonic transport has been observed at zero magnetic field when $v_{drift} = s$, the feature was explained by a mechanism different from PIROs \cite{hatke2015SAND,raichev2017}.}\BibitemShut {Stop}%
\bibitem [{\citenamefont {Mani}\ \emph {et~al.}(2013)\citenamefont {Mani}, \citenamefont {Kriisa},\ and\ \citenamefont {Wegscheider}}]{mani2013size}%
  \BibitemOpen
  \bibfield  {author} {\bibinfo {author} {\bibfnamefont {R.~G.}\ \bibnamefont {Mani}}, \bibinfo {author} {\bibfnamefont {A.}~\bibnamefont {Kriisa}},\ and\ \bibinfo {author} {\bibfnamefont {W.}~\bibnamefont {Wegscheider}},\ }\href {https://doi.org/10.1038/srep02747} {\bibfield  {journal} {\bibinfo  {journal} {Sci. Rep.}\ }\textbf {\bibinfo {volume} {3}},\ \bibinfo {pages} {2747} (\bibinfo {year} {2013})}\BibitemShut {NoStop}%
\bibitem [{\citenamefont {Bockhorn}\ \emph {et~al.}(2011)\citenamefont {Bockhorn}, \citenamefont {Barthold}, \citenamefont {Schuh}, \citenamefont {Wegscheider},\ and\ \citenamefont {Haug}}]{Bockhorn2011}%
  \BibitemOpen
  \bibfield  {author} {\bibinfo {author} {\bibfnamefont {L.}~\bibnamefont {Bockhorn}}, \bibinfo {author} {\bibfnamefont {P.}~\bibnamefont {Barthold}}, \bibinfo {author} {\bibfnamefont {D.}~\bibnamefont {Schuh}}, \bibinfo {author} {\bibfnamefont {W.}~\bibnamefont {Wegscheider}},\ and\ \bibinfo {author} {\bibfnamefont {R.~J.}\ \bibnamefont {Haug}},\ }\href {https://doi.org/10.1103/PhysRevB.83.113301} {\bibfield  {journal} {\bibinfo  {journal} {Phys. Rev. B}\ }\textbf {\bibinfo {volume} {83}},\ \bibinfo {pages} {113301} (\bibinfo {year} {2011})}\BibitemShut {NoStop}%
\bibitem [{\citenamefont {Bockhorn}\ \emph {et~al.}(2013)\citenamefont {Bockhorn}, \citenamefont {Hodaei}, \citenamefont {Schuh}, \citenamefont {Wegscheider},\ and\ \citenamefont {Haug}}]{Bockhorn2013}%
  \BibitemOpen
  \bibfield  {author} {\bibinfo {author} {\bibfnamefont {L.}~\bibnamefont {Bockhorn}}, \bibinfo {author} {\bibfnamefont {A.}~\bibnamefont {Hodaei}}, \bibinfo {author} {\bibfnamefont {D.}~\bibnamefont {Schuh}}, \bibinfo {author} {\bibfnamefont {W.}~\bibnamefont {Wegscheider}},\ and\ \bibinfo {author} {\bibfnamefont {R.~J.}\ \bibnamefont {Haug}},\ }\href {https://doi.org/10.1088/1742-6596/456/1/012003} {\bibfield  {journal} {\bibinfo  {journal} {J. Phys. Conf. Ser.}\ }\textbf {\bibinfo {volume} {456}},\ \bibinfo {pages} {012003} (\bibinfo {year} {2013})}\BibitemShut {NoStop}%
\bibitem [{\citenamefont {Shi}\ \emph {et~al.}(2014{\natexlab{a}})\citenamefont {Shi}, \citenamefont {Martin}, \citenamefont {Ebner}, \citenamefont {Zudov}, \citenamefont {Pfeiffer},\ and\ \citenamefont {West}}]{shi2014}%
  \BibitemOpen
  \bibfield  {author} {\bibinfo {author} {\bibfnamefont {Q.}~\bibnamefont {Shi}}, \bibinfo {author} {\bibfnamefont {P.~D.}\ \bibnamefont {Martin}}, \bibinfo {author} {\bibfnamefont {Q.~A.}\ \bibnamefont {Ebner}}, \bibinfo {author} {\bibfnamefont {M.~A.}\ \bibnamefont {Zudov}}, \bibinfo {author} {\bibfnamefont {L.~N.}\ \bibnamefont {Pfeiffer}},\ and\ \bibinfo {author} {\bibfnamefont {K.~W.}\ \bibnamefont {West}},\ }\href {https://doi.org/10.1103/PhysRevB.89.201301} {\bibfield  {journal} {\bibinfo  {journal} {Phys. Rev. B}\ }\textbf {\bibinfo {volume} {89}},\ \bibinfo {pages} {201301(R)} (\bibinfo {year} {2014}{\natexlab{a}})}\BibitemShut {NoStop}%
\bibitem [{\citenamefont {Shi}\ \emph {et~al.}(2014{\natexlab{b}})\citenamefont {Shi}, \citenamefont {Zudov}, \citenamefont {Pfeiffer},\ and\ \citenamefont {West}}]{Shi2014c}%
  \BibitemOpen
  \bibfield  {author} {\bibinfo {author} {\bibfnamefont {Q.}~\bibnamefont {Shi}}, \bibinfo {author} {\bibfnamefont {M.}~\bibnamefont {Zudov}}, \bibinfo {author} {\bibfnamefont {L.}~\bibnamefont {Pfeiffer}},\ and\ \bibinfo {author} {\bibfnamefont {K.}~\bibnamefont {West}},\ }\href@noop {} {\bibfield  {journal} {\bibinfo  {journal} {Phys. Rev. B}\ }\textbf {\bibinfo {volume} {90}},\ \bibinfo {pages} {201301(R)} (\bibinfo {year} {2014}{\natexlab{b}})}\BibitemShut {NoStop}%
\bibitem [{\citenamefont {Bockhorn}\ \emph {et~al.}(2014)\citenamefont {Bockhorn}, \citenamefont {Gornyi}, \citenamefont {Schuh}, \citenamefont {Reichl}, \citenamefont {Wegscheider},\ and\ \citenamefont {Haug}}]{Bockhorn2014}%
  \BibitemOpen
  \bibfield  {author} {\bibinfo {author} {\bibfnamefont {L.}~\bibnamefont {Bockhorn}}, \bibinfo {author} {\bibfnamefont {I.~V.}\ \bibnamefont {Gornyi}}, \bibinfo {author} {\bibfnamefont {D.}~\bibnamefont {Schuh}}, \bibinfo {author} {\bibfnamefont {C.}~\bibnamefont {Reichl}}, \bibinfo {author} {\bibfnamefont {W.}~\bibnamefont {Wegscheider}},\ and\ \bibinfo {author} {\bibfnamefont {R.~J.}\ \bibnamefont {Haug}},\ }\href {https://doi.org/10.1103/PhysRevB.90.165434} {\bibfield  {journal} {\bibinfo  {journal} {Phys. Rev. B}\ }\textbf {\bibinfo {volume} {90}},\ \bibinfo {pages} {165434} (\bibinfo {year} {2014})}\BibitemShut {NoStop}%
\bibitem [{\citenamefont {Schluck}\ \emph {et~al.}(2015)\citenamefont {Schluck}, \citenamefont {Fasbender}, \citenamefont {Heinzel}, \citenamefont {Pierz}, \citenamefont {Schumacher}, \citenamefont {Kazazis},\ and\ \citenamefont {Gennser}}]{Schluck2015}%
  \BibitemOpen
  \bibfield  {author} {\bibinfo {author} {\bibfnamefont {J.}~\bibnamefont {Schluck}}, \bibinfo {author} {\bibfnamefont {S.}~\bibnamefont {Fasbender}}, \bibinfo {author} {\bibfnamefont {T.}~\bibnamefont {Heinzel}}, \bibinfo {author} {\bibfnamefont {K.}~\bibnamefont {Pierz}}, \bibinfo {author} {\bibfnamefont {H.~W.}\ \bibnamefont {Schumacher}}, \bibinfo {author} {\bibfnamefont {D.}~\bibnamefont {Kazazis}},\ and\ \bibinfo {author} {\bibfnamefont {U.}~\bibnamefont {Gennser}},\ }\href {https://doi.org/10.1103/PhysRevB.91.195303} {\bibfield  {journal} {\bibinfo  {journal} {Phys. Rev. B}\ }\textbf {\bibinfo {volume} {91}},\ \bibinfo {pages} {195303} (\bibinfo {year} {2015})}\BibitemShut {NoStop}%
\bibitem [{\citenamefont {Wang}\ \emph {et~al.}(2016)\citenamefont {Wang}, \citenamefont {Samaraweera}, \citenamefont {Reichl}, \citenamefont {Wegscheider},\ and\ \citenamefont {Mani}}]{Wang2016}%
  \BibitemOpen
  \bibfield  {author} {\bibinfo {author} {\bibfnamefont {Z.}~\bibnamefont {Wang}}, \bibinfo {author} {\bibfnamefont {R.~L.}\ \bibnamefont {Samaraweera}}, \bibinfo {author} {\bibfnamefont {C.}~\bibnamefont {Reichl}}, \bibinfo {author} {\bibfnamefont {W.}~\bibnamefont {Wegscheider}},\ and\ \bibinfo {author} {\bibfnamefont {R.~G.}\ \bibnamefont {Mani}},\ }\href {https://doi.org/10.1038/srep38516} {\bibfield  {journal} {\bibinfo  {journal} {Sci. Rep.}\ }\textbf {\bibinfo {volume} {6}},\ \bibinfo {pages} {38516} (\bibinfo {year} {2016})}\BibitemShut {NoStop}%
\bibitem [{\citenamefont {Samaraweera}\ \emph {et~al.}(2017)\citenamefont {Samaraweera}, \citenamefont {Liu}, \citenamefont {Wang}, \citenamefont {Reichl}, \citenamefont {Wegscheider},\ and\ \citenamefont {Mani}}]{Samar2017}%
  \BibitemOpen
  \bibfield  {author} {\bibinfo {author} {\bibfnamefont {R.~L.}\ \bibnamefont {Samaraweera}}, \bibinfo {author} {\bibfnamefont {H.-C.}\ \bibnamefont {Liu}}, \bibinfo {author} {\bibfnamefont {Z.}~\bibnamefont {Wang}}, \bibinfo {author} {\bibfnamefont {C.}~\bibnamefont {Reichl}}, \bibinfo {author} {\bibfnamefont {W.}~\bibnamefont {Wegscheider}},\ and\ \bibinfo {author} {\bibfnamefont {R.~G.}\ \bibnamefont {Mani}},\ }\href {https://doi.org/10.1038/s41598-017-05351-8} {\bibfield  {journal} {\bibinfo  {journal} {Sci. Rep.}\ }\textbf {\bibinfo {volume} {7}},\ \bibinfo {pages} {5074} (\bibinfo {year} {2017})}\BibitemShut {NoStop}%
\bibitem [{\citenamefont {Samaraweera}\ \emph {et~al.}(2018)\citenamefont {Samaraweera}, \citenamefont {Liu}, \citenamefont {Gunawardana}, \citenamefont {Kriisa}, \citenamefont {Reichl}, \citenamefont {Wegscheider},\ and\ \citenamefont {Mani}}]{Samar2018}%
  \BibitemOpen
  \bibfield  {author} {\bibinfo {author} {\bibfnamefont {R.~L.}\ \bibnamefont {Samaraweera}}, \bibinfo {author} {\bibfnamefont {H.-C.}\ \bibnamefont {Liu}}, \bibinfo {author} {\bibfnamefont {B.}~\bibnamefont {Gunawardana}}, \bibinfo {author} {\bibfnamefont {A.}~\bibnamefont {Kriisa}}, \bibinfo {author} {\bibfnamefont {C.}~\bibnamefont {Reichl}}, \bibinfo {author} {\bibfnamefont {W.}~\bibnamefont {Wegscheider}},\ and\ \bibinfo {author} {\bibfnamefont {R.~G.}\ \bibnamefont {Mani}},\ }\href {https://doi.org/10.1038/s41598-018-28359-0} {\bibfield  {journal} {\bibinfo  {journal} {Sci. Rep.}\ }\textbf {\bibinfo {volume} {8}},\ \bibinfo {pages} {10061} (\bibinfo {year} {2018})}\BibitemShut {NoStop}%
\bibitem [{\citenamefont {Schluck}\ \emph {et~al.}(2018)\citenamefont {Schluck}, \citenamefont {Hund}, \citenamefont {Heckenthaler}, \citenamefont {Heinzel}, \citenamefont {Siboni}, \citenamefont {Horbach}, \citenamefont {Pierz}, \citenamefont {Schumacher}, \citenamefont {Kazazis}, \citenamefont {Gennser},\ and\ \citenamefont {Mailly}}]{Schluck2018}%
  \BibitemOpen
  \bibfield  {author} {\bibinfo {author} {\bibfnamefont {J.}~\bibnamefont {Schluck}}, \bibinfo {author} {\bibfnamefont {M.}~\bibnamefont {Hund}}, \bibinfo {author} {\bibfnamefont {T.}~\bibnamefont {Heckenthaler}}, \bibinfo {author} {\bibfnamefont {T.}~\bibnamefont {Heinzel}}, \bibinfo {author} {\bibfnamefont {N.~H.}\ \bibnamefont {Siboni}}, \bibinfo {author} {\bibfnamefont {J.}~\bibnamefont {Horbach}}, \bibinfo {author} {\bibfnamefont {K.}~\bibnamefont {Pierz}}, \bibinfo {author} {\bibfnamefont {H.~W.}\ \bibnamefont {Schumacher}}, \bibinfo {author} {\bibfnamefont {D.}~\bibnamefont {Kazazis}}, \bibinfo {author} {\bibfnamefont {U.}~\bibnamefont {Gennser}},\ and\ \bibinfo {author} {\bibfnamefont {D.}~\bibnamefont {Mailly}},\ }\href {https://doi.org/10.1103/PhysRevB.97.115301} {\bibfield  {journal} {\bibinfo  {journal} {Phys. Rev. B}\ }\textbf {\bibinfo {volume} {97}},\ \bibinfo {pages} {115301} (\bibinfo {year} {2018})}\BibitemShut {NoStop}%
\bibitem [{\citenamefont {Samaraweera}\ \emph {et~al.}(2020)\citenamefont {Samaraweera}, \citenamefont {Gunawardana}, \citenamefont {Nanayakkara}, \citenamefont {Munasinghe}, \citenamefont {Kriisa}, \citenamefont {Reichl}, \citenamefont {Wegscheider},\ and\ \citenamefont {Mani}}]{Samar2020}%
  \BibitemOpen
  \bibfield  {author} {\bibinfo {author} {\bibfnamefont {R.~L.}\ \bibnamefont {Samaraweera}}, \bibinfo {author} {\bibfnamefont {B.}~\bibnamefont {Gunawardana}}, \bibinfo {author} {\bibfnamefont {T.~R.}\ \bibnamefont {Nanayakkara}}, \bibinfo {author} {\bibfnamefont {R.~C.}\ \bibnamefont {Munasinghe}}, \bibinfo {author} {\bibfnamefont {A.}~\bibnamefont {Kriisa}}, \bibinfo {author} {\bibfnamefont {C.}~\bibnamefont {Reichl}}, \bibinfo {author} {\bibfnamefont {W.}~\bibnamefont {Wegscheider}},\ and\ \bibinfo {author} {\bibfnamefont {R.~G.}\ \bibnamefont {Mani}},\ }\href {https://doi.org/10.1038/s41598-019-57331-9} {\bibfield  {journal} {\bibinfo  {journal} {Sci. Rep.}\ }\textbf {\bibinfo {volume} {10}},\ \bibinfo {pages} {781} (\bibinfo {year} {2020})}\BibitemShut {NoStop}%
\bibitem [{\citenamefont {Horn-Cosfeld}\ \emph {et~al.}(2021)\citenamefont {Horn-Cosfeld}, \citenamefont {Schluck}, \citenamefont {Lammert}, \citenamefont {Cerchez}, \citenamefont {Heinzel}, \citenamefont {Pierz}, \citenamefont {Schumacher},\ and\ \citenamefont {Mailly}}]{HornCosfeld2021}%
  \BibitemOpen
  \bibfield  {author} {\bibinfo {author} {\bibfnamefont {B.}~\bibnamefont {Horn-Cosfeld}}, \bibinfo {author} {\bibfnamefont {J.}~\bibnamefont {Schluck}}, \bibinfo {author} {\bibfnamefont {J.}~\bibnamefont {Lammert}}, \bibinfo {author} {\bibfnamefont {M.}~\bibnamefont {Cerchez}}, \bibinfo {author} {\bibfnamefont {T.}~\bibnamefont {Heinzel}}, \bibinfo {author} {\bibfnamefont {K.}~\bibnamefont {Pierz}}, \bibinfo {author} {\bibfnamefont {H.~W.}\ \bibnamefont {Schumacher}},\ and\ \bibinfo {author} {\bibfnamefont {D.}~\bibnamefont {Mailly}},\ }\href {https://doi.org/10.1103/PhysRevB.104.045306} {\bibfield  {journal} {\bibinfo  {journal} {Phys. Rev. B}\ }\textbf {\bibinfo {volume} {104}},\ \bibinfo {pages} {045306} (\bibinfo {year} {2021})}\BibitemShut {NoStop}%
\bibitem [{\citenamefont {Wang}\ \emph {et~al.}(2022)\citenamefont {Wang}, \citenamefont {Jia}, \citenamefont {Du}, \citenamefont {Pfeiffer}, \citenamefont {Baldwin},\ and\ \citenamefont {West}}]{Wang2022}%
  \BibitemOpen
  \bibfield  {author} {\bibinfo {author} {\bibfnamefont {X.}~\bibnamefont {Wang}}, \bibinfo {author} {\bibfnamefont {P.}~\bibnamefont {Jia}}, \bibinfo {author} {\bibfnamefont {R.-R.}\ \bibnamefont {Du}}, \bibinfo {author} {\bibfnamefont {L.~N.}\ \bibnamefont {Pfeiffer}}, \bibinfo {author} {\bibfnamefont {K.~W.}\ \bibnamefont {Baldwin}},\ and\ \bibinfo {author} {\bibfnamefont {K.~W.}\ \bibnamefont {West}},\ }\href {https://doi.org/10.1103/PhysRevB.106.L241302} {\bibfield  {journal} {\bibinfo  {journal} {Phys. Rev. B}\ }\textbf {\bibinfo {volume} {106}},\ \bibinfo {pages} {L241302} (\bibinfo {year} {2022})}\BibitemShut {NoStop}%
\bibitem [{\citenamefont {Wang}\ \emph {et~al.}(2023)\citenamefont {Wang}, \citenamefont {Hilke}, \citenamefont {Fong}, \citenamefont {Austing}, \citenamefont {Studenikin}, \citenamefont {West},\ and\ \citenamefont {Pfeiffer}}]{paper1}%
  \BibitemOpen
  \bibfield  {author} {\bibinfo {author} {\bibfnamefont {Z.~T.}\ \bibnamefont {Wang}}, \bibinfo {author} {\bibfnamefont {M.}~\bibnamefont {Hilke}}, \bibinfo {author} {\bibfnamefont {N.}~\bibnamefont {Fong}}, \bibinfo {author} {\bibfnamefont {D.~G.}\ \bibnamefont {Austing}}, \bibinfo {author} {\bibfnamefont {S.~A.}\ \bibnamefont {Studenikin}}, \bibinfo {author} {\bibfnamefont {K.~W.}\ \bibnamefont {West}},\ and\ \bibinfo {author} {\bibfnamefont {L.~N.}\ \bibnamefont {Pfeiffer}},\ }\href {https://doi.org/10.1103/PhysRevB.107.195406} {\bibfield  {journal} {\bibinfo  {journal} {Phys. Rev. B}\ }\textbf {\bibinfo {volume} {107}},\ \bibinfo {pages} {195406} (\bibinfo {year} {2023})}\BibitemShut {NoStop}%
\bibitem [{\citenamefont {Bockhorn}\ \emph {et~al.}(2024)\citenamefont {Bockhorn}, \citenamefont {Schuh}, \citenamefont {Reichl}, \citenamefont {Wegscheider},\ and\ \citenamefont {Haug}}]{Bockhorn2024}%
  \BibitemOpen
  \bibfield  {author} {\bibinfo {author} {\bibfnamefont {L.}~\bibnamefont {Bockhorn}}, \bibinfo {author} {\bibfnamefont {D.}~\bibnamefont {Schuh}}, \bibinfo {author} {\bibfnamefont {C.}~\bibnamefont {Reichl}}, \bibinfo {author} {\bibfnamefont {W.}~\bibnamefont {Wegscheider}},\ and\ \bibinfo {author} {\bibfnamefont {R.~J.}\ \bibnamefont {Haug}},\ }\href {https://doi.org/10.1103/PhysRevB.109.205416} {\bibfield  {journal} {\bibinfo  {journal} {Phys. Rev. B}\ }\textbf {\bibinfo {volume} {109}},\ \bibinfo {pages} {205416} (\bibinfo {year} {2024})}\BibitemShut {NoStop}%
\bibitem [{\citenamefont {Bartels}\ \emph {et~al.}(2025)\citenamefont {Bartels}, \citenamefont {Strobel}, \citenamefont {Horn-Cosfeld}, \citenamefont {Cerchez}, \citenamefont {Pierz}, \citenamefont {Schumacher}, \citenamefont {Mailly},\ and\ \citenamefont {Heinzel}}]{Bartels2025}%
  \BibitemOpen
  \bibfield  {author} {\bibinfo {author} {\bibfnamefont {F.}~\bibnamefont {Bartels}}, \bibinfo {author} {\bibfnamefont {J.}~\bibnamefont {Strobel}}, \bibinfo {author} {\bibfnamefont {B.}~\bibnamefont {Horn-Cosfeld}}, \bibinfo {author} {\bibfnamefont {M.}~\bibnamefont {Cerchez}}, \bibinfo {author} {\bibfnamefont {K.}~\bibnamefont {Pierz}}, \bibinfo {author} {\bibfnamefont {H.~W.}\ \bibnamefont {Schumacher}}, \bibinfo {author} {\bibfnamefont {D.}~\bibnamefont {Mailly}},\ and\ \bibinfo {author} {\bibfnamefont {T.}~\bibnamefont {Heinzel}},\ }\href {https://doi.org/10.1103/PhysRevB.111.165301} {\bibfield  {journal} {\bibinfo  {journal} {Phys. Rev. B}\ }\textbf {\bibinfo {volume} {111}},\ \bibinfo {pages} {165301} (\bibinfo {year} {2025})}\BibitemShut {NoStop}%
\bibitem [{\citenamefont {Yang}\ \emph {et~al.}(2002{\natexlab{b}})\citenamefont {Yang}, \citenamefont {Zhang}, \citenamefont {Du}, \citenamefont {Simmons},\ and\ \citenamefont {Reno}}]{yang2002zener}%
  \BibitemOpen
  \bibfield  {author} {\bibinfo {author} {\bibfnamefont {C.~L.}\ \bibnamefont {Yang}}, \bibinfo {author} {\bibfnamefont {J.}~\bibnamefont {Zhang}}, \bibinfo {author} {\bibfnamefont {R.~R.}\ \bibnamefont {Du}}, \bibinfo {author} {\bibfnamefont {J.~A.}\ \bibnamefont {Simmons}},\ and\ \bibinfo {author} {\bibfnamefont {J.~L.}\ \bibnamefont {Reno}},\ }\href {https://doi.org/10.1103/PhysRevLett.89.076801} {\bibfield  {journal} {\bibinfo  {journal} {Phys. Rev. Lett.}\ }\textbf {\bibinfo {volume} {89}},\ \bibinfo {pages} {076801} (\bibinfo {year} {2002}{\natexlab{b}})}\BibitemShut {NoStop}%
\bibitem [{\citenamefont {Bykov}\ \emph {et~al.}(2005{\natexlab{b}})\citenamefont {Bykov}, \citenamefont {Zhang}, \citenamefont {Vitkalov}, \citenamefont {Kalagin},\ and\ \citenamefont {Bakarov}}]{Bykov2005}%
  \BibitemOpen
  \bibfield  {author} {\bibinfo {author} {\bibfnamefont {A.~A.}\ \bibnamefont {Bykov}}, \bibinfo {author} {\bibfnamefont {J.-q.}\ \bibnamefont {Zhang}}, \bibinfo {author} {\bibfnamefont {S.}~\bibnamefont {Vitkalov}}, \bibinfo {author} {\bibfnamefont {A.~K.}\ \bibnamefont {Kalagin}},\ and\ \bibinfo {author} {\bibfnamefont {A.~K.}\ \bibnamefont {Bakarov}},\ }\href {https://doi.org/10.1103/PhysRevB.72.245307} {\bibfield  {journal} {\bibinfo  {journal} {Phys. Rev. B}\ }\textbf {\bibinfo {volume} {72}},\ \bibinfo {pages} {245307} (\bibinfo {year} {2005}{\natexlab{b}})}\BibitemShut {NoStop}%
\bibitem [{\citenamefont {Zhang}\ \emph {et~al.}(2007{\natexlab{a}})\citenamefont {Zhang}, \citenamefont {Chiang}, \citenamefont {Zudov}, \citenamefont {Pfeiffer},\ and\ \citenamefont {West}}]{zhang2007MT}%
  \BibitemOpen
  \bibfield  {author} {\bibinfo {author} {\bibfnamefont {W.}~\bibnamefont {Zhang}}, \bibinfo {author} {\bibfnamefont {H.-S.}\ \bibnamefont {Chiang}}, \bibinfo {author} {\bibfnamefont {M.}~\bibnamefont {Zudov}}, \bibinfo {author} {\bibfnamefont {L.}~\bibnamefont {Pfeiffer}},\ and\ \bibinfo {author} {\bibfnamefont {K.}~\bibnamefont {West}},\ }\href {https://doi.org/10.1103/PhysRevB.75.041304} {\bibfield  {journal} {\bibinfo  {journal} {Phys. Rev. B}\ }\textbf {\bibinfo {volume} {75}},\ \bibinfo {pages} {041304(R)} (\bibinfo {year} {2007}{\natexlab{a}})}\BibitemShut {NoStop}%
\bibitem [{\citenamefont {Zhang}\ \emph {et~al.}(2007{\natexlab{b}})\citenamefont {Zhang}, \citenamefont {Vitkalov}, \citenamefont {Bykov}, \citenamefont {Kalagin},\ and\ \citenamefont {Bakarov}}]{zhang2007effect}%
  \BibitemOpen
  \bibfield  {author} {\bibinfo {author} {\bibfnamefont {J.-q.}\ \bibnamefont {Zhang}}, \bibinfo {author} {\bibfnamefont {S.}~\bibnamefont {Vitkalov}}, \bibinfo {author} {\bibfnamefont {A.}~\bibnamefont {Bykov}}, \bibinfo {author} {\bibfnamefont {A.}~\bibnamefont {Kalagin}},\ and\ \bibinfo {author} {\bibfnamefont {A.}~\bibnamefont {Bakarov}},\ }\href {https://doi.org/10.1103/PhysRevB.75.081305} {\bibfield  {journal} {\bibinfo  {journal} {Phys. Rev. B}\ }\textbf {\bibinfo {volume} {75}},\ \bibinfo {pages} {081305(R)} (\bibinfo {year} {2007}{\natexlab{b}})}\BibitemShut {NoStop}%
\bibitem [{\citenamefont {Vavilov}\ \emph {et~al.}(2007)\citenamefont {Vavilov}, \citenamefont {Aleiner},\ and\ \citenamefont {Glazman}}]{vavilov2007}%
  \BibitemOpen
  \bibfield  {author} {\bibinfo {author} {\bibfnamefont {M.~G.}\ \bibnamefont {Vavilov}}, \bibinfo {author} {\bibfnamefont {I.~L.}\ \bibnamefont {Aleiner}},\ and\ \bibinfo {author} {\bibfnamefont {L.~I.}\ \bibnamefont {Glazman}},\ }\href {https://doi.org/10.1103/PhysRevB.76.115331} {\bibfield  {journal} {\bibinfo  {journal} {Phys. Rev. B}\ }\textbf {\bibinfo {volume} {76}},\ \bibinfo {pages} {115331} (\bibinfo {year} {2007})}\BibitemShut {NoStop}%
\bibitem [{\citenamefont {Lei}(2007)}]{Lei2007}%
  \BibitemOpen
  \bibfield  {author} {\bibinfo {author} {\bibfnamefont {X.~L.}\ \bibnamefont {Lei}},\ }\href {https://doi.org/10.1063/1.2717572} {\bibfield  {journal} {\bibinfo  {journal} {Appl. Phys. Lett.}\ }\textbf {\bibinfo {volume} {90}},\ \bibinfo {pages} {132119} (\bibinfo {year} {2007})}\BibitemShut {NoStop}%
\bibitem [{\citenamefont {Bykov}(2008)}]{bykov2008}%
  \BibitemOpen
  \bibfield  {author} {\bibinfo {author} {\bibfnamefont {A.~A.}\ \bibnamefont {Bykov}},\ }\href {https://doi.org/10.1134/S0021364008180112} {\bibfield  {journal} {\bibinfo  {journal} {Sov. Phys. JETP}\ }\textbf {\bibinfo {volume} {88}},\ \bibinfo {pages} {394} (\bibinfo {year} {2008})}\BibitemShut {NoStop}%
\bibitem [{\citenamefont {Hatke}\ \emph {et~al.}(2009{\natexlab{b}})\citenamefont {Hatke}, \citenamefont {Zudov}, \citenamefont {Pfeiffer},\ and\ \citenamefont {West}}]{Hatke2009}%
  \BibitemOpen
  \bibfield  {author} {\bibinfo {author} {\bibfnamefont {A.~T.}\ \bibnamefont {Hatke}}, \bibinfo {author} {\bibfnamefont {M.~A.}\ \bibnamefont {Zudov}}, \bibinfo {author} {\bibfnamefont {L.~N.}\ \bibnamefont {Pfeiffer}},\ and\ \bibinfo {author} {\bibfnamefont {K.~W.}\ \bibnamefont {West}},\ }\href {https://doi.org/10.1103/PhysRevB.79.161308} {\bibfield  {journal} {\bibinfo  {journal} {Phys. Rev. B}\ }\textbf {\bibinfo {volume} {79}},\ \bibinfo {pages} {161308(R)} (\bibinfo {year} {2009}{\natexlab{b}})}\BibitemShut {NoStop}%
\bibitem [{\citenamefont {Hatke}\ \emph {et~al.}(2011{\natexlab{b}})\citenamefont {Hatke}, \citenamefont {Zudov}, \citenamefont {Pfeiffer},\ and\ \citenamefont {West}}]{Hatke2011}%
  \BibitemOpen
  \bibfield  {author} {\bibinfo {author} {\bibfnamefont {A.~T.}\ \bibnamefont {Hatke}}, \bibinfo {author} {\bibfnamefont {M.~A.}\ \bibnamefont {Zudov}}, \bibinfo {author} {\bibfnamefont {L.~N.}\ \bibnamefont {Pfeiffer}},\ and\ \bibinfo {author} {\bibfnamefont {K.~W.}\ \bibnamefont {West}},\ }\href {https://doi.org/10.1103/PhysRevB.83.081301} {\bibfield  {journal} {\bibinfo  {journal} {Phys. Rev. B}\ }\textbf {\bibinfo {volume} {83}},\ \bibinfo {pages} {081301(R)} (\bibinfo {year} {2011}{\natexlab{b}})}\BibitemShut {NoStop}%
\bibitem [{\citenamefont {Hatke}\ \emph {et~al.}(2012{\natexlab{b}})\citenamefont {Hatke}, \citenamefont {Zudov}, \citenamefont {Pfeiffer},\ and\ \citenamefont {West}}]{Hatke2012}%
  \BibitemOpen
  \bibfield  {author} {\bibinfo {author} {\bibfnamefont {A.~T.}\ \bibnamefont {Hatke}}, \bibinfo {author} {\bibfnamefont {M.~A.}\ \bibnamefont {Zudov}}, \bibinfo {author} {\bibfnamefont {L.~N.}\ \bibnamefont {Pfeiffer}},\ and\ \bibinfo {author} {\bibfnamefont {K.~W.}\ \bibnamefont {West}},\ }\href {https://doi.org/10.1103/PhysRevB.86.081307} {\bibfield  {journal} {\bibinfo  {journal} {Phys. Rev. B}\ }\textbf {\bibinfo {volume} {86}},\ \bibinfo {pages} {081307(R)} (\bibinfo {year} {2012}{\natexlab{b}})}\BibitemShut {NoStop}%
\bibitem [{\citenamefont {Shi}\ \emph {et~al.}(2014{\natexlab{c}})\citenamefont {Shi}, \citenamefont {Ebner},\ and\ \citenamefont {Zudov}}]{Shi2014b}%
  \BibitemOpen
  \bibfield  {author} {\bibinfo {author} {\bibfnamefont {Q.}~\bibnamefont {Shi}}, \bibinfo {author} {\bibfnamefont {Q.~A.}\ \bibnamefont {Ebner}},\ and\ \bibinfo {author} {\bibfnamefont {M.~A.}\ \bibnamefont {Zudov}},\ }\href {https://doi.org/10.1103/PhysRevB.90.161301} {\bibfield  {journal} {\bibinfo  {journal} {Phys. Rev. B}\ }\textbf {\bibinfo {volume} {90}},\ \bibinfo {pages} {161301(R)} (\bibinfo {year} {2014}{\natexlab{c}})}\BibitemShut {NoStop}%
\bibitem [{\citenamefont {Shi}\ \emph {et~al.}(2017)\citenamefont {Shi}, \citenamefont {Zudov}, \citenamefont {Falson}, \citenamefont {Kozuka}, \citenamefont {Tsukazaki}, \citenamefont {Kawasaki}, \citenamefont {von Klitzing},\ and\ \citenamefont {Smet}}]{shi2017}%
  \BibitemOpen
  \bibfield  {author} {\bibinfo {author} {\bibfnamefont {Q.}~\bibnamefont {Shi}}, \bibinfo {author} {\bibfnamefont {M.~A.}\ \bibnamefont {Zudov}}, \bibinfo {author} {\bibfnamefont {J.}~\bibnamefont {Falson}}, \bibinfo {author} {\bibfnamefont {Y.}~\bibnamefont {Kozuka}}, \bibinfo {author} {\bibfnamefont {A.}~\bibnamefont {Tsukazaki}}, \bibinfo {author} {\bibfnamefont {M.}~\bibnamefont {Kawasaki}}, \bibinfo {author} {\bibfnamefont {K.}~\bibnamefont {von Klitzing}},\ and\ \bibinfo {author} {\bibfnamefont {J.}~\bibnamefont {Smet}},\ }\href {https://doi.org/10.1103/PhysRevB.95.041411} {\bibfield  {journal} {\bibinfo  {journal} {Phys. Rev. B}\ }\textbf {\bibinfo {volume} {95}},\ \bibinfo {pages} {041411(R)} (\bibinfo {year} {2017})}\BibitemShut {NoStop}%
\bibitem [{\citenamefont {Zudov}\ \emph {et~al.}(2017)\citenamefont {Zudov}, \citenamefont {Dmitriev}, \citenamefont {Friess}, \citenamefont {Shi}, \citenamefont {Umansky}, \citenamefont {von Klitzing},\ and\ \citenamefont {Smet}}]{Zudov2017}%
  \BibitemOpen
  \bibfield  {author} {\bibinfo {author} {\bibfnamefont {M.~A.}\ \bibnamefont {Zudov}}, \bibinfo {author} {\bibfnamefont {I.~A.}\ \bibnamefont {Dmitriev}}, \bibinfo {author} {\bibfnamefont {B.}~\bibnamefont {Friess}}, \bibinfo {author} {\bibfnamefont {Q.}~\bibnamefont {Shi}}, \bibinfo {author} {\bibfnamefont {V.}~\bibnamefont {Umansky}}, \bibinfo {author} {\bibfnamefont {K.}~\bibnamefont {von Klitzing}},\ and\ \bibinfo {author} {\bibfnamefont {J.}~\bibnamefont {Smet}},\ }\href {https://doi.org/10.1103/PhysRevB.96.121301} {\bibfield  {journal} {\bibinfo  {journal} {Phys. Rev. B}\ }\textbf {\bibinfo {volume} {96}},\ \bibinfo {pages} {121301(R)} (\bibinfo {year} {2017})}\BibitemShut {NoStop}%
\bibitem [{\citenamefont {Yu}\ \emph {et~al.}(2018)\citenamefont {Yu}, \citenamefont {Hilke}, \citenamefont {Poole}, \citenamefont {Studenikin},\ and\ \citenamefont {Austing}}]{phaseinversionYu}%
  \BibitemOpen
  \bibfield  {author} {\bibinfo {author} {\bibfnamefont {V.}~\bibnamefont {Yu}}, \bibinfo {author} {\bibfnamefont {M.}~\bibnamefont {Hilke}}, \bibinfo {author} {\bibfnamefont {P.~J.}\ \bibnamefont {Poole}}, \bibinfo {author} {\bibfnamefont {S.}~\bibnamefont {Studenikin}},\ and\ \bibinfo {author} {\bibfnamefont {D.~G.}\ \bibnamefont {Austing}},\ }\href {https://doi.org/10.1103/PhysRevB.98.165434} {\bibfield  {journal} {\bibinfo  {journal} {Phys. Rev. B}\ }\textbf {\bibinfo {volume} {98}},\ \bibinfo {pages} {165434} (\bibinfo {year} {2018})}\BibitemShut {NoStop}%
\bibitem [{\citenamefont {Ponomarev}\ and\ \citenamefont {Efros}(2001)}]{Ponomarev2001}%
  \BibitemOpen
  \bibfield  {author} {\bibinfo {author} {\bibfnamefont {I.~V.}\ \bibnamefont {Ponomarev}}\ and\ \bibinfo {author} {\bibfnamefont {A.~L.}\ \bibnamefont {Efros}},\ }\href {https://doi.org/10.1103/PhysRevB.63.165305} {\bibfield  {journal} {\bibinfo  {journal} {Phys. Rev. B}\ }\textbf {\bibinfo {volume} {63}},\ \bibinfo {pages} {165305} (\bibinfo {year} {2001})}\BibitemShut {NoStop}%
\bibitem [{\citenamefont {Ryzhii}(2003)}]{Ryzhii2003}%
  \BibitemOpen
  \bibfield  {author} {\bibinfo {author} {\bibfnamefont {V.}~\bibnamefont {Ryzhii}},\ }\href {https://doi.org/10.1103/PhysRevB.68.193402} {\bibfield  {journal} {\bibinfo  {journal} {Phys. Rev. B}\ }\textbf {\bibinfo {volume} {68}},\ \bibinfo {pages} {193402} (\bibinfo {year} {2003})}\BibitemShut {NoStop}%
\bibitem [{\citenamefont {Lei}(2008)}]{Lei2008}%
  \BibitemOpen
  \bibfield  {author} {\bibinfo {author} {\bibfnamefont {X.~L.}\ \bibnamefont {Lei}},\ }\href {https://doi.org/10.1103/PhysRevB.77.205309} {\bibfield  {journal} {\bibinfo  {journal} {Phys. Rev. B}\ }\textbf {\bibinfo {volume} {77}},\ \bibinfo {pages} {205309} (\bibinfo {year} {2008})}\BibitemShut {NoStop}%
\bibitem [{\citenamefont {Raichev}(2009)}]{raichev2009}%
  \BibitemOpen
  \bibfield  {author} {\bibinfo {author} {\bibfnamefont {O.~E.}\ \bibnamefont {Raichev}},\ }\href {https://doi.org/10.1103/PhysRevB.80.075318} {\bibfield  {journal} {\bibinfo  {journal} {Phys. Rev. B}\ }\textbf {\bibinfo {volume} {80}},\ \bibinfo {pages} {075318} (\bibinfo {year} {2009})}\BibitemShut {NoStop}%
\bibitem [{\citenamefont {Keser}\ \emph {et~al.}(2021)\citenamefont {Keser}, \citenamefont {Wang}, \citenamefont {Klochan}, \citenamefont {Ho}, \citenamefont {Tkachenko}, \citenamefont {Tkachenko}, \citenamefont {Culcer}, \citenamefont {Adam}, \citenamefont {Farrer}, \citenamefont {Ritchie}, \citenamefont {Sushkov},\ and\ \citenamefont {Hamilton}}]{Keser2021}%
  \BibitemOpen
  \bibfield  {author} {\bibinfo {author} {\bibfnamefont {A.~C.}\ \bibnamefont {Keser}}, \bibinfo {author} {\bibfnamefont {D.~Q.}\ \bibnamefont {Wang}}, \bibinfo {author} {\bibfnamefont {O.}~\bibnamefont {Klochan}}, \bibinfo {author} {\bibfnamefont {D.~Y.~H.}\ \bibnamefont {Ho}}, \bibinfo {author} {\bibfnamefont {O.~A.}\ \bibnamefont {Tkachenko}}, \bibinfo {author} {\bibfnamefont {V.~A.}\ \bibnamefont {Tkachenko}}, \bibinfo {author} {\bibfnamefont {D.}~\bibnamefont {Culcer}}, \bibinfo {author} {\bibfnamefont {S.}~\bibnamefont {Adam}}, \bibinfo {author} {\bibfnamefont {I.}~\bibnamefont {Farrer}}, \bibinfo {author} {\bibfnamefont {D.~A.}\ \bibnamefont {Ritchie}}, \bibinfo {author} {\bibfnamefont {O.~P.}\ \bibnamefont {Sushkov}},\ and\ \bibinfo {author} {\bibfnamefont {A.~R.}\ \bibnamefont {Hamilton}},\ }\href {https://doi.org/10.1103/PhysRevX.11.031030} {\bibfield  {journal} {\bibinfo  {journal} {Phys. Rev. X}\ }\textbf {\bibinfo {volume} {11}},\ \bibinfo {pages} {031030} (\bibinfo {year} {2021})}\BibitemShut
  {NoStop}%
\bibitem [{\citenamefont {Ashlea~Alava}\ \emph {et~al.}(2024)\citenamefont {Ashlea~Alava}, \citenamefont {Kumar}, \citenamefont {Harsas}, \citenamefont {Mehta}, \citenamefont {Hathi}, \citenamefont {Chen}, \citenamefont {Ritchie},\ and\ \citenamefont {Hamilton}}]{Ashlea2024}%
  \BibitemOpen
  \bibfield  {author} {\bibinfo {author} {\bibfnamefont {Y.}~\bibnamefont {Ashlea~Alava}}, \bibinfo {author} {\bibfnamefont {K.}~\bibnamefont {Kumar}}, \bibinfo {author} {\bibfnamefont {C.}~\bibnamefont {Harsas}}, \bibinfo {author} {\bibfnamefont {P.}~\bibnamefont {Mehta}}, \bibinfo {author} {\bibfnamefont {P.}~\bibnamefont {Hathi}}, \bibinfo {author} {\bibfnamefont {C.}~\bibnamefont {Chen}}, \bibinfo {author} {\bibfnamefont {D.~A.}\ \bibnamefont {Ritchie}},\ and\ \bibinfo {author} {\bibfnamefont {A.~R.}\ \bibnamefont {Hamilton}},\ }\href {https://doi.org/10.1063/5.0234082} {\bibfield  {journal} {\bibinfo  {journal} {Appl. Phys. Lett.}\ }\textbf {\bibinfo {volume} {125}},\ \bibinfo {pages} {252105} (\bibinfo {year} {2024})}\BibitemShut {NoStop}%
\end{thebibliography}%
